\documentclass[prd,twocolumn,superscriptaddress,floatfix,amsmath,amssymb,amsfonts,longbibliography,nofootinbib]{revtex4-1}

\usepackage{float} 

\usepackage[normalem]{ulem}
\usepackage[english]{babel}
\usepackage{graphicx}
\usepackage{dcolumn}
\usepackage{bm}
\usepackage{blindtext}
\usepackage{verbatim}
\usepackage{relsize}
\usepackage{mathrsfs}
\usepackage{musicography}
\usepackage{amsmath}
\usepackage{blindtext}
\usepackage{cancel}
\usepackage{physics}
\usepackage{epstopdf}
\usepackage{mathtools}
\usepackage{blindtext}
\usepackage{tensor}
\usepackage{color}
\usepackage[usenames,dvipsnames]{pstricks}
\usepackage{epsfig}
\usepackage{pst-grad} 
\usepackage{pst-plot} 
\usepackage{hyperref}
\usepackage{verbatim}
\usepackage{slashed}
\usepackage{dsfont}

\newcommand{\mf}{\mathsf}

\newcommand{\ii}{\mathrm{i}}
\newcommand{\tb}[1]{\textcolor{black}{#1}}

\allowdisplaybreaks[1] 

\begin{document}

\title{Localized non-relativistic quantum systems in curved spacetimes: \\
a general characterization of particle detector models}

\author{T. Rick Perche}
\email{trickperche@perimeterinstitute.ca}

\affiliation{Department of Applied Mathematics, University of Waterloo, Waterloo, Ontario, N2L 3G1, Canada}
\affiliation{Perimeter Institute for Theoretical Physics, Waterloo, Ontario, N2L 2Y5, Canada}
\affiliation{Institute for Quantum Computing, University of Waterloo, Waterloo, Ontario, N2L 3G1, Canada}

\begin{abstract}
    In this manuscript we provide a consistent way of describing a localized non-relativistic quantum system undergoing a timelike trajectory in a background curved spacetime. Namely, using Fermi normal coordinates, we identify an inner product and canonically conjugate position and momentum operators defined in the rest space of the trajectory for each value of its proper time. This framework then naturally provides a recipe for mapping a quantum theory defined in a non-relativistic background to a theory around a timelike trajectory in curved spacetimes. This is done by reinterpreting the position and momentum operators and by introducing a local redshift factor to the Hamiltonian, which gives rise to new dynamics due to the curvature of spacetime and the acceleration of the trajectory. We then apply our formalism to particle detector models, that is, to the case where the non-relativistic quantum system is coupled to a quantum field in a curved background. This allows one to write a general definition for particle detector models which is able to recover the previous models in the literature. Our framework also allows one to estimate the regime of validity of these models, characterizing the situations where particle detectors can be used to accurately probe quantum fields.
\end{abstract}


\maketitle

\section{Introduction}

The interface between gravity and quantum mechanics has been among the hottest topics of theoretical physics for the past decades. As of yet, we don't currently have a widely agreed upon theory that provides a quantum description for gravity. However, classical general relativity can accommodate quantum systems in numerous regimes. One example is the description of quantum fields in curved spacetimes using a semiclassical approximation. This approach yields fundamental results for theoretical physics, such as the Unruh effect~\cite{fullingUnruhEffect,Unruh-Wald,Unruh1976,Schlicht,waitUnruh}, Hawking radiation~\cite{HawkingRadiation,Davies1974,HawkingGibons} and the inflationary description of the early universe~\cite{inflation}. Another example of growing recent interest is the description of non-relativistic quantum systems in curved spacetimes. Part of this interest emerges from the field of relativistic quantum information (RQI), which considers localized probes in order to explore information tasks in quantum field theory both in flat and curved spacetimes.

Non-relativistic quantum mechanics is not the fundamental theory for the description of the world. This is not only due to its inconsistencies with special and general relativity, but also because it is known to be the limit of a more fundamental theory, which is necessary to describe experiments we have access to. However, most of the world that is accessible to us is well described by non-relativistic systems. The atoms that constitute our day-to-day life, and most quantum experiments that we have access to are very well described by non-relativistic quantum mechanics. Moreover, while our current understanding of quantum field theory is enough to describe most interactions perturbatively, and yields very accurate predictions for scattering processes and decay rates, the description of bound states in quantum field theory is far from being fully understood. For instance, we currently do not have a fully satisfactory description of a Hydrogen atom in terms of quantum electrodynamics. Overall, be it due to convenience, or ignorance, the description of many phenomena in the universe requires us to treat systems using the formalism of non-relativistic quantum mechanics.

The main goal of this manuscript is to provide a clear and consistent formulation for a localized non-relativistic quantum system undergoing a timelike trajectory in a background curved spacetime. We do so by employing Fermi normal coordinates around the trajectory, which define rest spaces $\Sigma_\tau$, where the system's wavefunction can be defined. Then, we find canonically conjugate position and momentum operators, which allow one to describe a non-relativistic quantum system in curved spacetimes. We adapt the system's Hamiltonian to curved spacetimes by adding a redshift factor, which matches previous descriptions found in the literature. This Hamiltonian then defines a unitary time-evolution operator which evolves the system's state between rest spaces. We then find bounds for the regime of validity of our theory. In essence, we find that it is valid whenever the size of the system is sufficiently smaller compared to the curvature radius of spacetime, and to the inverse of the trajectory's proper acceleration. We also find new dynamics to the system's internal degrees of freedom due to this description, and explore its consequences in different setups. 

We then apply this framework to describe particle detector models, usually considered in the field of RQI. These models have been introduced in~\cite{Unruh1976,DeWitt} and have been extensively used in the literature of RQI to describe many information protocols in quantum field theory, such as entanglement harvesting~\cite{Valentini1991,Reznik1,vacuumEntanglement,vacuumBell,Petar,Pozas-Kerstjens:2015,Pozas2016,ericksonBH,ericksonNew,carol}, entanglement farming~\cite{Farming}, communication protocols via quantum fields~\cite{Benincasa_2014,Jonsson2,Jonsson3,Landulfo,martin-martinez2015,Jonsson4,Katja,Simidzija_2020}, and quantum energy teleportation~\cite{teleportation,nichoTeleport}. In general, a particle detector model is a localized quantum system with non-relativistic degrees of freedom that couples to a quantum field. These models are attractive both from a theoretical and from a practical perspective, since they have been shown to describe the interaction of atoms with light~\cite{Pozas2016,Nicho1,richard}, and the interaction of nucleons with the neutrino fields~\cite{neutrinos,antiparticles,carol}. But particle detectors are more than this: they are the general physical tools that allow one to obtain local information about quantum field theories. In fact, particle detector models have been used to explore the Unruh~\cite{Unruh1976,Takagi,matsasUnruh,waitUnruh,unruhEffectNoThermal,unruhSlow,garay,mine} and Hawking effects~\cite{Unruh1976,bhDetectors,bhDetectorsAdS,bhDetectorsBTZ}, accurately recover the two-point functions of quantum field theories~\cite{pipo,geometry}, recover the geometry of spacetime~\cite{geometry,ahmed}, and develop a consistent measurement theory for quantum fields~\cite{chicken}. 

Our description of localized non-relativistic quantum systems then allows us to define a general notion of particle detector model, and to precisely quantify the regimes where these models can be used to accurately probe quantum field theories. We also find that the detector's acceleration and the curvature of spacetime change its internal dynamics, effectively adding noise to their outcomes, which impacts their ability to probe quantum fields. Our framework then allows one to precisely quantify this noise and to impose a regime of validity for the application of particle detector models.

This manuscript is organized as follows. In Section \ref{sec:FNC} we review the Fermi normal coordinates around a worldline and define the $\emph{Fermi bound}$, which will be used to describe non-relativistic quantum systems in their proper frame and estimate the regime of validity of our description. In Section \ref{sec:NRQS} we develop the framework which allows one to consistently describe a general non-relativistic quantum system with a position degree of freedom in curved spacetimes, and study the new dynamics introduced by the system's motion and the curvature of spacetime. In Section \ref{sec:detectors} we apply the framework of Section \ref{sec:NRQS} to particle detector models, defining a general notion of particle detector model, and deriving its limit of validity. The conclusions of this work can be found in Section \ref{sec:conclusions}.

\section{Fermi Normal Coordinates and the Fermi bound}\label{sec:FNC}

This section has two main goals: to review Fermi normal coordinates, so that we can introduce the notation used throughout the manuscript, and to define the Fermi bound. We study these in Subsections \ref{sub:fnc} and \ref{sub:fermibound}, respectively.

\subsection{Fermi normal coordinates}\label{sub:fnc}

Let $\mathcal{M}$ denote an $n+1$ dimensional spacetime with Lorentzian metric $g$, and consider a timelike trajectory $\mf z(\tau)$ in $\mathcal{M}$, parametrized by its proper time \mbox{$\tau\in(\tau_{\text{min}},\tau_{\text{max}})$}\footnote{The formalism also allows for the case where $\tau_{\text{min}} = -\infty$ and $\tau_{\text{max}} = \infty$.}. The Fermi normal coordinates around the trajectory $\mf z(\tau)$ are coordinates which are able to describe physically relevant quantities associated to an observer undergoing the trajectory. The Fermi normal coordinates are also useful because one can expand the metric components in a neighbourhood of the curve in terms of the curvature of spacetime and the trajectory's proper acceleration.

The time component of the Fermi normal coordinates is defined as the proper time of the curve, $\tau$. In order to define the spacelike coordinates, $\bm x$, we first pick an orthonormal frame $\mf{e}_\mu(\tau_0)$ in the tangent space to a given point of the curve, $T_{\mf z(\tau_0)}\mathcal{M}$ such that $e_0^\mu(\tau_0) = u^\mu(\tau_0)$ is the four-velocity of the curve. Then, we have
\begin{equation}
    g(\mf e_\mu,\mf e_\nu) = \eta_{\mu\nu},
\end{equation}
where $\eta_{\mu\nu} = \text{diag}(-1,1,1,1)$. The next step is to extend this frame along the curve $\mf z(\tau)$. In order to do this, we transport the vectors $\mf e_\mu(\tau_0)$ via the Fermi transport:
\begin{equation}
    \frac{\text{D}(e_\mu)^\alpha}{\dd \tau} + 2a^{[\alpha}u^{\beta]}(e_\mu)_\beta = 0, 
\end{equation}
where $\frac{\text{D}}{\dd \tau}$ denotes the covariant derivative along $\mf z(\tau)$ and $a^\mu = \frac{\text{D}u^\mu}{\dd \tau}$ is the proper acceleration of the trajectory. The Fermi transport takes into account the natural motion of the curve in order to transport vectors between different tangent spaces. Notice that because $u_\mu a^\mu = 0$, the four-velocity is always Fermi transported along the curve. Thus, Fermi transporting the frame $\mf e_\mu(\tau_0)$ along $\mf z(\tau)$ gives a frame $\mf e_\mu(\tau)$, such that $\mf e^\mu_0 = u^\mu$ for all $\tau$. This frame will be referred to as the Fermi frame.

We define the spacelike Fermi normal coordinates \mbox{$\bm x=(x^1,...,x^n)$} as follows. Let $\mathcal{N}_\mf{p}$ denote the normal neighbourhood of $\mf p$, that is, the set of all points which can be connected to $\mf p$ by a unique geodesic. For a given $\tau$, we define the rest surface $\Sigma_\tau\subset\mathcal{N}_{\mf z(\tau)}$ as the set spanned by all geodesics starting at $\mf z(\tau)$ with tangent vector orthogonal to $u^\mu$. The surfaces $\Sigma_\tau$ correspond to the local rest spaces around $\mf z (\tau)$ and define a local foliation of spacetime around the curve. Let $\mf p\in\Sigma_\tau$ for some $\tau$, then we assign coordinates $(\tau,x^1,...,x^n)$ to $\mf p$ if $\mf p = \text{exp}_{\mf z(\tau)}(x^i\mf e_i(\tau))$, where $\exp_{\mf z(\tau)}$ denotes the exponential map at the point $\mf z(\tau)$. The Fermi normal coordinates are well defined in the world tube $\mathcal{F} = \bigcup_\tau \Sigma_\tau$ around the trajectory, so that any point $\mf x\in \mathcal{F}$ can be identified as $\mf x = (\tau,\bm x)$. A consequence of the definition is that the proper distance of a point $\mf x$ to the curve $\mf z(\tau)$ is given by $r = \sqrt{\delta_{ij}x^ix^j}$, so that proper distances from $\mf z(\tau)$ can be computed using the Euclidean norm of the spacelike Fermi normal coordinates. 

It is important to mention that although the time parameter of the Fermi normal coordinates is the proper time of the trajectory $\mf z (\tau)$, in general it does not correspond to the proper time of the other trajectories defined by $\bm x = \text{const.}$ In fact, in a general curved spacetime the vector $\partial_\tau$ is not normal to the surfaces $\Sigma_\tau$, and not normalized at $\bm x \neq 0$.

It is also useful to define a local orthonormal frame associated to the Fermi normal coordinates by extending the Fermi frame to the tube $\mathcal{F}$. For a given event \mbox{$\mf x \in \tb{\mathcal{F}}$}, we define the \emph{extended Fermi frame} $\mf e_\mu(\mf x)$ by parallel transporting the vectors $\mf e_\mu(\tau)$ along the geodesic contained in $\Sigma_\tau$ that connects $\mf z(\tau)$ to $\mf x$. This process then defines an orthonormal frame in every point within the region $\mathcal{F}$.

It is also possible to find an expression for the metric components in Fermi normal coordinates in terms of an expansion on the physical distance of a point to the curve, $r=\sqrt{\delta_{ij}x^ix^j}$. The expansion reads
\begin{align}\label{eq:expansionFNC}
        &g_{\tau \tau}=-\left(1+a_i(\tau) x^i\right)^2-R_{0 {{i}} 0 {{j}}}(\tau)  x^{i}  x^{j} + \mathcal{O}(r^3),\nonumber\\
        &g_{\tau i}=-\frac{2}{3} R_{0 {{j}}{{i}}{{k}}}(\tau)  x^{j}  x^{k}+ \mathcal{O}(r^3),\nonumber\\
        &g_{ij}=\delta_{{{i}}{{j}}}-\frac{1}{3} R_{{{i}}{{k}}{{j}}{{l}}}(\tau)  x^{k}  x^{l}+ \mathcal{O}(r^3),
\end{align}
where $a_\mu(\tau)$ and $R_{\mu\nu\alpha\beta}(\tau)$ denote the components of acceleration and curvature in Fermi coordinates at $\mf z(\tau)$. This expansion is valid if $|\bm x|$ is sufficiently smaller than both the curvature radius of spacetime and $1/a$, where $a = \sqrt{a^\mu a_\mu}$ is the magnitude of the proper acceleration of the curve. The expansion of Eq. \eqref{eq:expansionFNC} has found many uses in the literature, such as providing a treatment for extended bodies in general relativity~\cite{DixonI,DixonII,DixonIII}, finding the energy level shift on a hydrogen atom due to curvature~\cite{parker,ParkerRevLett,ParkerHydrogen}, describing the motion of point charges in curved spacetimes~\cite{poisson}, and, more recently, describing localized non-relativistic systems in curved spacetimes~\cite{us,jonas,mine,theguy}.

\subsection{The Fermi bound}\label{sub:fermibound}

In this subsection we define, estimate, and discuss a quantity with units of length, which we name the \emph{Fermi bound}. The Fermi bound is essentially the maximum radius that a system centered at the curve $\mf z(\tau)$ can have in order to be completely described in terms of Fermi normal coordinates.

\begin{figure}[h!]
    \centering
    \includegraphics[width=8cm]{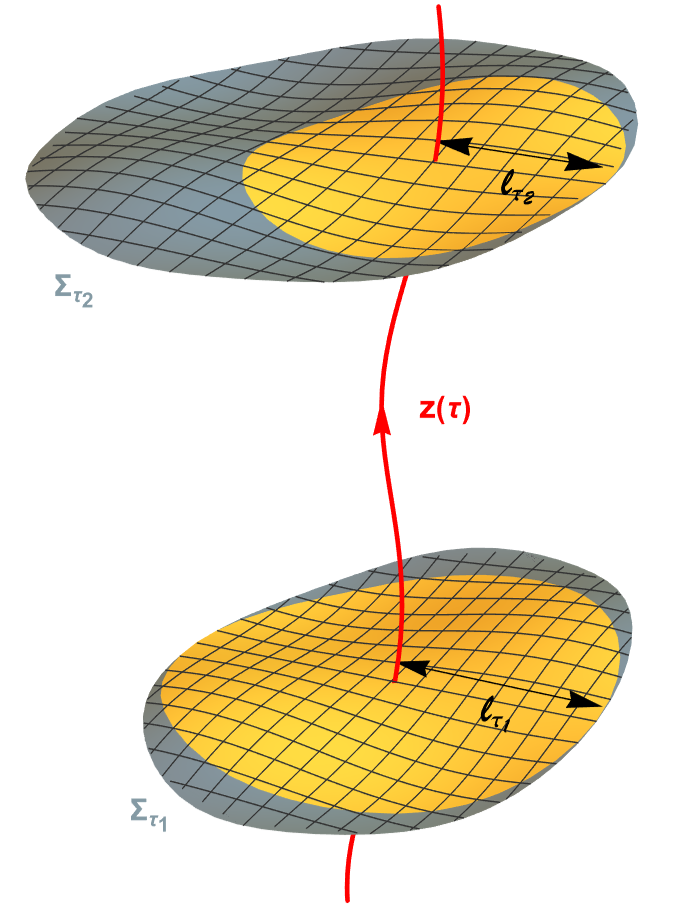}
    \caption{Schematic representation of the region delimited by the $\tau$-Fermi bound (in yellow) within each constant $\tau$ surface $\Sigma_\tau$ (in gray).}
    \label{fig:FermiScheme}
\end{figure}

We first define the $\tau$-Fermi bound. Consider the set of spacelike geodesics which connect $\mf z(\tau)$ to the boundary of $\Sigma_\tau$. The $\tau$-Fermi bound $\ell_\tau$ is defined as the minimum proper length of maximally extended geodesics in this set. In essence, it is the largest radius that a spacelike ball $B\subset T_{\mf z(\tau)}\mathcal{M}$ orthogonal to $u^\mu(\tau)$ can have so that $\exp_{\mf z(\tau)}(B)\subset\Sigma_\tau$. Thus, any system defined in $\Sigma_\tau$ which is centered at $\mf z(\tau)$ and contained in a ball with proper radius smaller than $\ell_\tau$ can be entirely described using Fermi normal coordinates. There are two parameters that control the size of $\ell_\tau$. The curve's acceleration effectively bends the surfaces $\Sigma_\tau$ so that some geodesics overlap after a length of $1/a$, even in flat spacetimes. Meanwhile, spacetime may be positively curved, which makes nearby geodesics converge, so that they overlap after a certain distance. Overall, $\ell_\tau$ is controlled by the curve's acceleration and the curvature of spacetime. A schematic representation of the region delimited by the $\tau$-Fermi bound within each rest space can be found in Fig. \ref{fig:FermiScheme}.

The \emph{Fermi bound} $\ell$ is defined as the infimum of the \mbox{$\tau$-Fermi} bounds. That is,
\begin{equation}
    \ell = \inf_\tau \,\,\ell_\tau.
\end{equation} 
Each of the $\tau$-Fermi bounds defines a bound for the size of a system in $\Sigma_\tau$ which can be described in terms of spacelike Fermi normal coordinates. Thus, the Fermi-bound is a bound for the size of a system centered at the curve $\mf z(\tau)$ which can be entirely described by Fermi normal coordinates at all times. The Fermi bound also defines a world tube around the trajectory, where systems that can be entirely described in Fermi normal coordinates may have support. This tube is defined as the region spanned by all geodesics contained in $\Sigma_\tau$ which have proper length smaller than the Fermi bound $\ell$ for each $\tau$.

\begin{figure}[h]
    \centering
    \includegraphics[width=8cm]{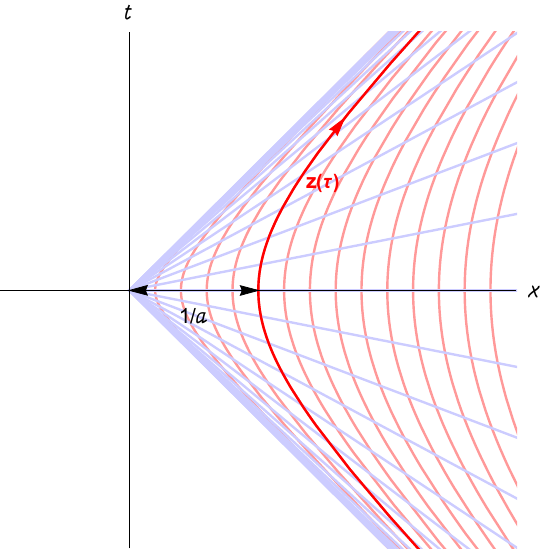}
    \caption{Fermi normal coordinates for a uniformly accelerated trajectory in Minkowski spacetime.}
    \label{fig:acceleration}
\end{figure}

An illustrative example of Fermi normal coordinates and the Fermi bound can be obtained for a uniformly accelerated trajectory in Minkowski spacetime. Consider inertial coordinates $(t,x,y,z)$ in Minkowski spacetime and a uniformly accelerated observer undergoing a trajectory $\mf z(\tau) = (\frac{1}{a}\cosh(a\tau),\frac{1}{a}\sinh(a\tau),0,0)$. Then, the Fermi normal coordinates around $\mf z(\tau)$ are the Rindler coordinates $(\tau,\bm x)$, with $\bm x = (X,y,z)$. The metric in these coordinates reads
\begin{equation}
    g = -(1+aX)^2 \dd\tau^2 + dX^2 + dy^2 + dz^2.
\end{equation}
From this expression one can see that the metric becomes degenerate at $X = -1/a$, which corresponds to the events of the form $(0,0,y,z)$ in inertial coordinates. This also corresponds to the events where the Fermi normal coordinates break down. Given that the proper distance is given by the Euclidean distance in the spacelike Fermi normal coordinates, we see that the Fermi bound for a uniformly accelerated trajectory is $\ell = 1/a$. The Fermi normal coordinates and the Fermi bound around a uniformly accelerated trajectory in Minkowski spacetime are depicted in Fig. \ref{fig:acceleration}. 

More generally, in Appendix \ref{app:fermi} we show that under the conditions where the expansion of Eq. \eqref{eq:expansionFNC} is valid, it is possible to estimate the Fermi bound from below by
\begin{equation}\label{eq:approximationFermi}
    \ell\gtrsim  \inf_\tau \left(\frac{1}{a+\sqrt{\lambda_R}}\right),
\end{equation}
where $a = \sqrt{a_\mu a^\mu}$ is the norm of the four acceleration of $\mf z(\tau)$ and $\lambda_R$ is the largest positive eigenvalue of the operator $-R_{0i0j}$, if there are any. This estimate can be useful for providing bounds for the regime of validity of frameworks which use Fermi normal coordinates. Notice that for the case of uniformly accelerated trajectories in Rindler space this estimate is exact.

\section{Non-relativistic quantum systems in curved spacetimes}\label{sec:NRQS}

In this section we will provide a framework that allows one to describe a localized non-relativistic quantum system in curved spacetimes. We will work with systems which can be described by a wavefunction in an $n$ dimensional space and by internal degrees of freedom defined in a finite dimensional space.

\subsection{A single particle in non-relativistic quantum mechanics}\label{sub:nonrel}

In order to obtain a generalization to curved spacetimes, let us first consider a system that can be described in terms of a wavefunction in a non-relativistic setup. We will assume that the system can be described in a Hilbert space $\mathcal{H} = \mathcal{H}_\textsc{x}\otimes\mathcal{H}_\textsc{s}$, where $\mathcal{H}_\textsc{x}$ is associated to the position degrees of freedom of the particle (its wavefunction), and $\mathcal{H}_\textsc{s}$ is a finite dimensional Hilbert space associated to its additional internal degrees of freedom (such as its spin). Then, the canonical variables associated to the position degrees of freedom in $\mathcal{H}_\textsc{x}$ are the position and momentum operators, $\hat{x}^i$ and $\hat{p}_j$. These satisfy the commutation relations
\begin{equation}\label{eq:xpcomm}
    \comm{\hat{x}^i}{\hat{p}_j} = \ii \delta^i_j\openone.
\end{equation}
In order to translate this description to curved spacetimes, it will be useful to work in the position representation of such system. Let $\ket{\bm x}$ denote the \tb{non-normalizable} eigenstates of $\hat{\bm x} = (\hat{x}^1,...,\hat{x}^n)$, and let $\ket{s}$ be any basis for $\mathcal{H}_\textsc{s}$. Then any state $\ket{\psi}$ can be written as
\begin{equation}
    \ket{\psi} = \sum_s \int \dd^n \bm x\, \braket{\bm x,s}{\psi}\,\ket{\bm x,s}.
\end{equation}
We define $\psi^s(\bm x) = \braket{\bm x,s}{\psi}$ as the wavefunction representation of $\ket{\psi}$ in the basis $\ket{s}$. Normalization of the state $\ket{\psi}$ then implies
\begin{align}
    \braket{\psi}{\psi} &= \sum_{s,s'} \int \dd^n \bm x\,\dd^n \bm x' \,(\psi^{s'}(\bm x'))^*\psi^s(\bm x) \braket{\bm x'}{\bm x}\braket{s'}{s}\nonumber\\
    &= \sum_{s,s'} \int \dd^n \bm x(\psi^{s'}(\bm x))^*\psi^s(\bm x) \delta_{s's}\nonumber\\
    &= \int \dd^n \bm x \,\psi_s^*(\bm x) \psi^s(\bm x)=1,
\end{align}
where we denote $\psi_s(\bm x)=\delta_{s's}\psi^{s'}(\bm x)$ and we used Einstein's summation convention from the second to third lines. That is, the components $\psi^s(\bm x)$ can be seen as elements of $L^2(\mathbb{R}^n)$. In fact, we have $\mathcal{H}_\textsc{x}\cong L^2(\mathbb{R}^n)$, where the isomorphism is $\ket{\psi}\longmapsto \psi(\bm x) = \braket{\bm x}{\psi}$. In the space $L^2(\mathbb{R}^n)$, the position operator acts in the wavefunctions as multiplication,
\begin{equation}\label{eq:x}
    \bra{\bm x}\hat{x}^i \ket{\psi} = x^i \psi(\bm x),
\end{equation}
and the momentum operator acts according to
\begin{equation}\label{eq:p}
    \bra{\bm x} \hat{p}_j \ket{\psi} = - \ii \partial_j \psi(\bm x).
\end{equation}
From Eqs. \eqref{eq:x} and \eqref{eq:p}, it is clear that the commutation relations of Eq. \eqref{eq:xpcomm} are satisfied. 

We assume that the dynamics of the system are prescribed by a self-adjoint Hamiltonian $\hat{H}(\hat{\bm x},\hat{\bm p}, \{\hat{s}_i\},t)$, where  $\{\hat{s}_i\}$ denotes any collection of operators acting in $\mathcal{H}_\textsc{s}$ and $t$ denotes a possible external time dependence on the Hamiltonian. This Hamiltonian then generates unitary time evolution according to Schr\"odinger's equation,
\begin{equation}
    \ii\dv{}{t}\ket{\psi(t)} = \hat{H}(t)\ket{\psi(t)},
\end{equation}
where we have omitted the dependence of $\hat{H}$ in $\hat{\bm x}$, $\hat{\bm p}$ and $\{\hat{s}_i\}$ to lighten the notation. Equivalently, one can write the time evolved state in terms of the time evolution operator $\hat{U}(t,t_0)$, defined by $\ket{\psi(t)} = \hat{U}(t,t_0)\ket{\psi(t_0)}$, so that Schr\"odinger's equation gives
\begin{equation}\label{eq:SchU}
    \ii\dv{}{t}\hat{U}(t,t_0) = \hat{H}(t)\hat{U}(t,t_0),
\end{equation}
which can be shown to define a unitary operator $\hat{U}(t,t_0)$. In fact, the solution to Eq. \eqref{eq:SchU} reads
\begin{equation}
    \hat{U}(t,t_0) = \mathcal{T}\text{exp}\left(-\ii \int \dd t \hat{H}(t)\right),
\end{equation}
where $\mathcal{T}\exp$ denotes the time ordered exponential. 

Finally, there is an important remark to be made regarding the system's Hamiltonian. Given that we will later provide a framework so that one could describe the system in a general relativistic setup, it will be important to consider the \emph{total} energy of the system, which takes into consideration its rest mass. This essentially amounts to adding a term $m c^2 \,\openone$ to the non-relativistic Hamiltonian\footnote{here $c$ denotes the speed of light, which we mostly set to $1$ throughout in the manuscript.}, where $m$ denotes the rest mass of the system. For instance, a particle of mass $m$ under the influence of a potential $V(\bm x)$ should be associated to the Hamiltonian
\begin{equation}\label{eq:HO}
    \hat{H} = m + \frac{\hat{\bm p}^2}{2m} + V(\hat{\bm x}).
\end{equation}
Notice that the introduction of the rest mass does not influence the dynamics of the system, given that it only amounts to an overall shift in the energy levels.


\subsection{A localized non-relativistic quantum system in curved spacetimes}\label{sub:main}

We now provide a framework for describing the localized quantum system from the previous subsection undergoing a timelike trajectory $\mf z(\tau)$ in a given $(n+1)$ dimensional background spacetime $\mathcal{M}$. We assume that the quantum system is localized in space at each instant of time. This corresponds to the assumption that there exists a timelike curve $\mf z(\tau)$, parametrizing the events around which the quantum system is localized. For convenience we will refer to $\mf z(\tau)$ as the trajectory of the system. We then assign Fermi normal coordinates $(\tau,\bm x)$ around the curve $\mf z(\tau)$, so that the local rest spaces of the system are the surfaces $\Sigma_\tau$, defined by constant values of the $\tau$ coordinate. Our goal here is to define the Hilbert space for the wavefunctions at each value of the time coordinate $\tau$ as $L^2(\Sigma_\tau)$ with a suitable integration measure. However, there are some important remarks that have to be considered. 

First, we notice that $\Sigma_\tau$ is only locally defined, and does not extend past the normal neighbourhood of $\mf z(\tau)$. This means that if one wishes to consider wavefunctions completely defined in $L^2(\Sigma_\tau)$, then one must consider functions which are defined in a finite sized box or, equivalently, that the potential which traps the system is infinite outside of a region centered at $\bm x = 0$ with radius smaller than the Fermi bound $\ell$. This condition can be relaxed if the potential is strong enough so that the effective localization of the system is mostly within a radius $\ell$ of the trajectory. Within this relaxed assumption, one loses information about the ``tails of the wavefunction''. Nevertheless, if these tails can be assumed negligible compared to the values of the wavefunction in the region where the surfaces $\Sigma_\tau$ are well defined, the description is approximately valid. Overall, we will call the assumption that the wavefunction is completely localized in each of the $\Sigma_\tau$ surfaces the assumption of \emph{Fermi localization}, and the assumption that the wavefunctions are approximately localized within the $\Sigma_\tau$ surfaces will be called \emph{approximate Fermi localization}. It is important to notice that a physical (finite) trapping potential cannot produce wavefunctions which are Fermi localized, but only approximately Fermi localized.

Second, it is important to mention that the formalism developed here is not a fundamental description, and cannot be valid for a system with arbitrarily high energies. In fact, an important assumption for our model is that the non-relativistic energy of the system is sufficiently small compared to its rest energy. This can also be formulated as the assumption that $\sqrt{\langle \hat{\bm p}^2\rangle}$ is small compared to the system's rest mass $m$, or, in other words, that the system's average velocity $\sqrt{\langle\hat{\bm v}^2\rangle}$ is small compared to the speed of light. As we will discuss, this assumption will ensure that the dynamics introduced by the motion of the system and the spacetime curvature reduce to corrections previously found in the literature in similar setups.

The first step in order to formulate our description for a non-relativistic system in curved spacetimes is to appropriately determine the inner product in $L^2(\Sigma_\tau)$. The natural choice is to define the inner product as the integral with respect to the measure of the surfaces $\Sigma_\tau$. That is, for $\psi(\bm x)$ and $\phi(\bm x)$ defined in $\Sigma_\tau$,
\begin{equation}\label{eq:innprod}
    (\psi,\phi)_\tau \equiv \int_{\Sigma_\tau} \dd \Sigma \, \psi^*(\bm x) \phi(\bm x),
\end{equation}
where $\dd \Sigma = \sqrt{g_\Sigma(\tau,\bm x)}\,\dd^n \bm x$ is the invariant volume measure in $\Sigma_\tau$, with $g_\Sigma = \det (h_{ij})$, and $h_{ij}$ denotes the induced metric in $\Sigma_\tau$. Here we see that the assumption of Fermi localization ensures that the wavefunctions above are well defined within the surface, and can be integrated in Eq. \eqref{eq:innprod}. Approximate Fermi localization then ensures that only the tails of the wavefunctions are neglected in Eq. \eqref{eq:innprod}. More than a geometrical and natural inner product, $(\psi,\phi)_\tau$ defined above is obtained when one considers the reduction of Dirac spinors to wavefunctions defined in local rest spaces, as was done in \cite{jonas}. We will also see that the inner product of Eq. \eqref{eq:innprod} allows one to define a consistent quantum theory, with self-adjoint canonically conjugate position and momentum operators for the system.

Under the assumption of Fermi localization, one can then find position and momentum operators defined in terms of their actions in wavefunctions $\psi(\bm x)\in L^2(\Sigma_\tau)$. We define the position operator $\hat{\bm x} = \hat{x}^i \mathsf{e}_i$, where $\mf{e}_i$ denotes the extended Fermi frame and the $\hat{x}^i$ are defined via their action in wavefunctions as
\begin{equation}\label{eq:posOp}
    \hat{x}^i:\psi(\bm x)\longmapsto x^i \psi(\bm x).
\end{equation}
That is, the generalization of each component of the position operator multiplies the wavefunction by the components of $\bm x$ in Fermi normal coordinates. Physically, the definition of Eq. \eqref{eq:posOp} is justified by the fact that in Fermi normal coordinates, $|\bm x|=\sqrt{\delta_{ij}x^ix^j}$ corresponds to the proper distance between a point and the center of the curve. It is also important to mention that the components of the position operator defined above are also self-adjoint with respect to the inner product defined in Eq. \eqref{eq:innprod}.

The momentum operator in the position representation can also be defined by its action on wavefunctions \mbox{$\psi(\bm x)\in\Sigma_\tau$}. We define its components by
\begin{equation}\label{eq:pOp}
    \hat{p}_j:\psi(\bm x) \longmapsto  \frac{- \ii}{(g_\Sigma)^\frac{1}{4}}\pdv{}{x^j} \left((g_\Sigma)^\frac{1}{4}\psi(\bm x)\right).
\end{equation}
Although the factors of $1/4$ may seem out of place at a first glance, they are necessary in order for the $\hat{p}_j$ operators to be self-adjoint with respect to the inner product of Eq. \eqref{eq:innprod}. In fact,
\begin{align}
    (\psi,\hat{p}_j\phi)_\tau &=  \int\dd^n\bm x \sqrt{g_{\Sigma}}\,\psi^*(\bm x) \frac{-\ii}{(g_\Sigma)^\frac{1}{4}}\partial_j \left((g_\Sigma)^\frac{1}{4}\phi(\bm x)\right)\nonumber\\
    &= -\ii \int \dd^n\bm x(g_\Sigma)^\frac{1}{4} \psi^*(\bm x) \partial_j \left((g_\Sigma)^\frac{1}{4}\phi(\bm x)\right)\nonumber\\
    &= \ii \int \dd^n\bm x\partial_j\left((g_\Sigma)^\frac{1}{4} \psi^*(\bm x)\right) (g_\Sigma)^\frac{1}{4}\phi(\bm x)\nonumber\\
    &= \int \dd^n\bm x\left(-\ii\partial_j\left((g_\Sigma)^\frac{1}{4} \psi\right)\right)^* (g_\Sigma)^\frac{1}{4}\phi(\bm x)\nonumber\\
    &= \int \dd^n\bm x\sqrt{g_{\Sigma}}\left(\frac{-\ii}{(g_\Sigma)^\frac{1}{4}}\partial_j\left((g_\Sigma)^\frac{1}{4} \psi(\bm x)\right)\right)^* \!\!\phi(\bm x) \nonumber\\
    &= (\hat{p}_j \psi,\phi)_\tau,
\end{align}
where we have integrated by parts in the third equality and the boundary terms vanish under the assumption of Fermi localization\footnote{Under the assumption of approximate Fermi localization, one would then obtain boundary terms of the same order as the terms lost in the inner product, yielding approximately self-adjoint $\hat{p}_j$ operators. In order to obtain a fully consistent framework, one can then effectively truncate approximately Fermi localized wavefunctions, so that their dynamics can be approximated by that of compactly supported wavefunctions.}. Not only are the momentum operator components defined in Eq. \eqref{eq:pOp} self-adjoint, they also satisfy the canonical commutation relations with the position operator of Eq. \eqref{eq:posOp}:
\begin{equation}
    \hat{x}^i \hat{p}_j - \hat{p}_j \hat{x}^i : \psi(\bm x) \longmapsto \ii \delta^i_j\psi(\bm x).
\end{equation}
Thus, the $\hat{x}^i$ and $\hat{p}_j$ operators defined here are a valid generalization of the position and momentum operators of quantum mechanics. In fact, given any non-relativistic quantum system described in terms of its position and momentum operators $\hat{\bm x}$ and $\hat{\bm p}$, one can describe it around a trajectory $\mf z(\tau)$ for a fixed $\tau$ by considering its wavefunction to be in $\Sigma_\tau$, and interpreting the corresponding position and momentum operators as Eqs. \eqref{eq:posOp} and \eqref{eq:pOp}. We remark that canonically conjugated position and momentum operators fully define a quantum theory for a wavefunction at a given time, so that the description provided here is indeed enough for describing the position degrees of freedom of the quantum state around the trajectory $\mf z(\tau)$.

In order to fully describe the system in curved spacetimes, one also requires to describe the additional internal degrees of freedom of the system contained in the collection of operators $\{\hat{s}_i\}$ in this more general setup. Given that these degrees of freedom are internal to the particle, we describe them in the same Hilbert space $\mathcal{H}_\textsc{s}$, with no modifications to the inner product when one goes to curved spacetimes. Notice, however, that although their description will not change, it might be necessary to introduce different dynamics for the $\{\hat{s}_i\}$ operators in order to describe their evolution in curved spacetimes\footnote{For instance, the dynamics of the spin of particles can be affected by spacetime curvature in a non-trivial way, as discussed in~\cite{parker,jonas,theguy}.}. This will be further discussed when we consider time evolution in this framework. 

The treatment given so far allows one to describe a non-relativistic quantum system in curved spacetimes at a given surface $\Sigma_\tau$. However, we have not yet mentioned how time evolution can be implemented in this description. In other words, we have yet to describe a Hamiltonian formulation in this setup. First, notice that at each value of the time parameter $\tau$, the wavefunctions are defined in a \emph{different} Hilbert space $L^2(\Sigma_\tau)$. This adds extra complications when writing Schr\"odinger's equation, as one cannot differentiate states with respect to time via a limit of infinitesimal differences, since $\psi(\tau_0+\delta\tau,\bm x)$ and $\psi(\tau_0,\bm x)$ are defined in different Hilbert spaces. In order to make sense of Schr\"odinger's equation in this setup, we must locally extend the wavefunctions defined in $\Sigma_{\tau_0}$, so that we obtain a function $\psi(\tau,\bm x)$ for $\tau\in[\tau_0,\tau_0 + \varepsilon)$ for a small $\varepsilon>0$. It is then possible to compare its values at different $\tau$'s, so that differentiation can be performed. This essentially amounts to differentiation of a scalar function locally defined in spacetime with respect to the time parameter $\tau$.

At this stage, one could naively think that given a Hamiltonian $\hat{H}(\hat{\bm x},\hat{\bm p},\{\hat{s}_i\},t)$ for a quantum particle in a non-relativistic setup, it is enough to replace its dependence on $\hat{\bm x}$, $\hat{\bm p}$ and $\{\hat{s}_i\}$ as described previously, together with the replacement $t\longmapsto\tau$ in order to write Schr\"odinger's equation. However, there is an important ingredient missing that also has to be considered: redshift. As mentioned in Section \ref{sec:FNC}, the time parameter $\tau$ only corresponds to the proper time of an observer along the curve $\mf z(\tau)$. This implies that the time evolution at each point of space should contain a redshift factor, associated to the time dilation of the foliation defined by the $\Sigma_\tau$ surfaces. In Appendix \ref{app:det}, we compute the corresponding redshift factor. It is given by
\begin{equation}\label{eq:redshift}
    \gamma(\tau,\bm x) = \abs{g_{\tau\tau} - g_{\tau i} g_{\tau j} h^{ij}}^{\frac{1}{2}},
\end{equation}
where $h^{ij}$ denotes the inverse of the induced metric in the $\Sigma_\tau$ surfaces. 

In a classical system, one would take this redshift factor into account by multiplying the local Hamiltonian of the system by $\gamma(\tau,{\bm x})$, giving rise to the effective Hamiltonian $\gamma(\tau,{\bm x})\hat{H}({\bm x},{\bm p},\{{s}_i\},\tau)$. In a quantum setup, one would then be tempted to describe the Hamiltonian as $\gamma(\tau,\hat{\bm x})\hat{H}(\hat{\bm x},\hat{\bm p},\{\hat{s}_i\},\tau)$, promoting the space dependence in $\gamma(\tau,\bm x)$ to the position operator $\hat{\bm x}$. However, this product will not necessarily be self-adjoint due to the dependence of $\hat{H}$ on $\hat{\bm p}$. In order to obtain a self-adjoint Hamiltonian, one could then use the Weyl quantization prescription~\cite{Weyl1927} for the Hamiltonian $\gamma(\tau,\hat{\bm x})\hat{H}(\hat{\bm x},\hat{\bm p},\{\hat{s}_i\},\tau)$ or use the Moyal product~\cite{moyal}. A simpler way to handle the self-adjointness problem is to define the Hamiltonian via a symmetrization as
\begin{equation}\label{eq:Hcurved}
    \hat{\mathcal{H}}(\hat{\bm x},\hat{\bm p},\{\hat{s}_i\},\tau) = \frac{1}{2}\left(\gamma(\tau,\hat{\bm x})\hat{H}(\hat{\bm x},\hat{\bm p},\{\hat{s}_i\},\tau) + \text{H.c.}\right).
\end{equation}
Although different quantization methods might in principle give different Hamiltonians, in Subsection \ref{sub:coincidence} we will argue that the Hamiltonian can be well approximated by \mbox{$\hat{H}(\hat{\bm x},\hat{\bm p},\{\hat{s}_i\},\tau)+ma_i(\tau)\hat{x}^i + \frac{m}{2}R_{0i0j}\hat{x}^i\hat{x}^j$}, where $m$ is the rest mass of the system. Moreover, this approximate correction is also independent of ordering ambiguities, and yields the same result for any quantization prescription chosen. 

We first write Schr\"odinger's equation for a system with no extra spin degrees of freedom. That is, in the case where the Hilbert space $\mathcal{H}_\textsc{s}$ is trivial, and the system can be entirely described by its wavefunction. In this case, Schr\"odinger's equation can be written simply as
\begin{equation}
    \ii\pdv{}{\tau}\psi(\tau,\bm x) =\hat{\mathcal{H}}(\hat{\bm x},\hat{\bm p},\tau)\psi(\tau,\bm x),
\end{equation}
where the $\hat{\bm x}$ and $\hat{\bm p}$ operators act in $\psi(\tau,\bm x)$ according to Eqs. \eqref{eq:posOp} and \eqref{eq:pOp}. We remark that although it might look like the newly introduced dynamics, and the extra factors in the differential operator $\hat{\bm p}$ give rise to a much more complicated differential equation, one can instead use the commutation relations between $\hat{\bm x}$ and $\hat{\bm p}$ in order to find the solutions to Schr\"odinger's equation (as is usually done with the quantum harmonic oscillator, for instance).

In order to write Schr\"odinger's equation when the system also has internal degrees of freedom in $\mathcal{H}_\textsc{s}$, we write states in the Dirac notation, with \mbox{$\ket{\psi}\in \mathcal{H}^{(\tau)}_{\textsc{x}}\otimes \mathcal{H}_\textsc{s}$}, where $\mathcal{H}^{(\tau)}_{\textsc{x}}\cong L^2(\Sigma_\tau)$ for each $\tau$. In this context, the position eigenvectors are $\ket{\bm x}$ such that $\psi^s(\tau,\bm x) = \braket{\bm x,s}{\psi(\tau)}$ and a decomposition of the identity in the position basis can be written as
\begin{equation}\label{eq:idDecomp}
    \openone = \sum_s \int \dd \Sigma \ket{\bm x,s}\!\!\bra{\bm x,s}.
\end{equation}
In terms of Dirac's notation, we can then write Schr\"odinger's equation as
\begin{equation}
    \ii \dv{}{\tau}\ket{\psi} = \hat{\mathcal{H}}(\tau) \ket{\psi},
\end{equation}
where the $\tau$ differentiation in the position spaces is understood via the local extension of the wavefunctions, and we omitted the dependence of $\hat{\mathcal{H}}$ in $\hat{\bm x}$, $\hat{\bm p}$ and $\{\hat{s}_i\}$ for simplicity. This equation also defines unitary operators $\hat{U}(\tau,\tau_0)$ by $\hat{U}(\tau,\tau_0)\ket{\psi(\tau_0)} = \ket{\psi(\tau)}$\footnote{It is important to keep in mind that this family of unitary operators acts in different Hilbert spaces. That is,
\begin{align}
    \hat{U}(\tau,\tau_0): \mathcal{H}^{(\tau_0)}_{\textsc{x}}\otimes \mathcal{H}_\textsc{s}&\longrightarrow  \mathcal{H}^{(\tau)}_{\textsc{x}}\otimes \mathcal{H}_\textsc{s}\\
    \:\:\:\:\ket{\psi(\tau_0)}&\longmapsto \ket{\psi(\tau)},\nonumber
\end{align}
where $\mathcal{H}^{(\tau_0)}_{\textsc{x}}$ and $\mathcal{H}^{(\tau)}_{\textsc{x}}$ denote the $L^2$ spaces at $\Sigma_{\tau_0}$ and $\Sigma_\tau$, respectively.}.

Finally, we comment on the possible need to perform additional changes to the Hamiltonian $\hat{H}$, apart from the redshift factor and the replacement of the position and momentum operators for their definitions of Eqs. \eqref{eq:posOp} and \eqref{eq:pOp}. These additional changes could come from interactions of the other internal degrees of freedom of the system (encoded in the $\{\hat{s}_i\}$ operators) with curvature and acceleration. Although it is not possible to give a general recipe for adapting general operators to curved spacetimes, the framework provided here can accommodate for these changes in each case with minor modifications. For instance, \tb{in~\cite{jonas,theguy}}, a fermionic particle in curved spacetimes is described, and the coupling of its spin with curvature is obtained. This could be implemented here by adding terms to the Hamiltonian of Eq. \eqref{eq:Hcurved} corresponding to this interaction.

\begin{figure}[h!]
    \centering
    \includegraphics[width=8cm]{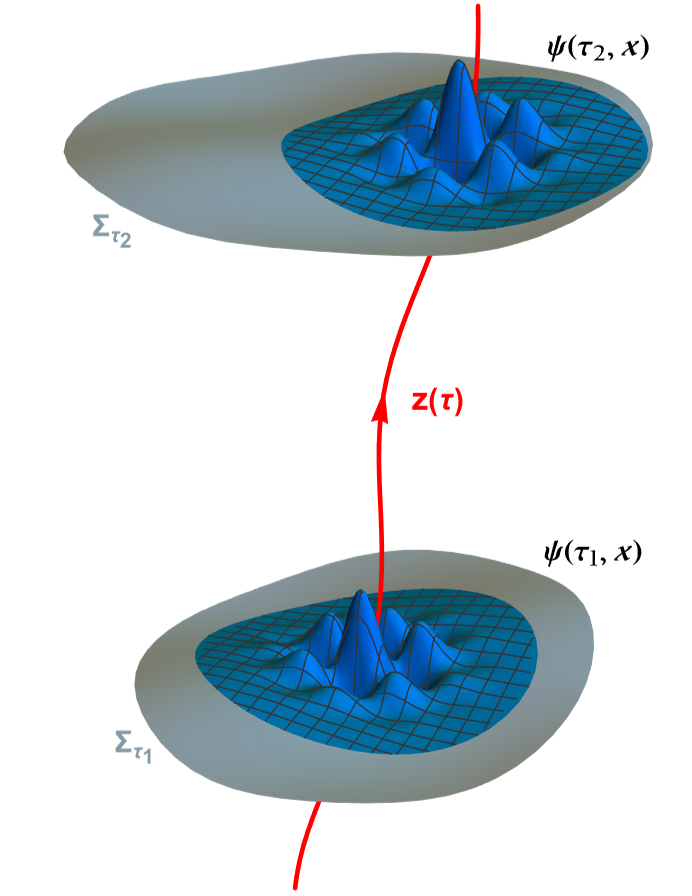}
    \caption{Schematic representation of the model for localized non-relativistic systems in curved spacetimes, with wavefunctions defined in the local rests paces.}
    \label{fig:WavefunctionScheme}
\end{figure}

Overall, in this section we completed one of the main goals of this manuscript: we provided a consistent description for a localized non-relativistic quantum system in curved spacetimes. A schematic representation of the model obtained can be found in Fig. \ref{fig:WavefunctionScheme}. Throughout the remaining of the manuscript we will discuss consequences and applications of this formalism.

\subsection{Discussion of the regime of validity of the model}\label{sub:regime}

In this section we discuss the regime of validity of the framework presented in Subsection \ref{sub:main}, and discuss the compatibility of the model with the framework of general relativity.

First, we remark once again that in order for the formalism presented in Subsection \ref{sub:main} to be applied, the non-relativistic quantum system must be sufficiently localized within the Fermi bound of the trajectory, $\ell$. For instance, the formulation works for systems which are compactly supported with support contained in a sphere of radius smaller than $\ell$. However, in order to perfectly trap a quantum particle, one requires an infinite potential, which is an unphysical assumption. If the potential is finite, but strong enough to approximately Fermi localize the system, the formalism of Subsection \ref{sub:main} can still be applied by neglecting the tails of the wavefunction outside of a sphere of radius $\ell$. Although one expects to lose some information about the system, the loss due to this approximation can be controlled by analyzing the tails of the wavefunctions associated with the relevant states of the system. 

It is possible to quantify the regime of validity of the theory using the estimate for the Fermi bound of Eq. \eqref{eq:approximationFermi}. Consider a non-relativistic quantum theory, which we wish to describe around a trajectory $\mf z(\tau)$ in curved spacetimes. We will work under the assumption that the system is \emph{strongly supported} within a region of radius $R(\varepsilon)$. The notion of strong support has been used by Eduardo Mart\'in-Mart\'inez to characterize functions that decay quickly with some characteristic scale since 2015~\cite{strongSupport}. In the context of our framework, a state of a quantum field will be said to be strongly supported within a region of radius $R(\varepsilon)$ if the expected value of a set of observables of interest can be computed to precision $\varepsilon$ by performing spatial integrals of the wavefunctions in a ball of radius $R(\varepsilon)$. The collection of observables of interest will explicitly depend on the system under consideration and on the observables that are relevant for the predictions one wishes to compute in each setup.

Within the assumption that the relevant states of the system have strong support within a region of radius $R(\varepsilon)$, the condition for our framework to be applicable is $R(\varepsilon)<\ell$, and the errors in the description are controlled by the parameter $\varepsilon$. Assuming the system to be sufficiently localized with respect to the curvature of spacetime and the trajectory's acceleration, one can then use the estimate of Eq. \eqref{eq:approximationFermi}, which gives the approximate condition
\begin{equation}
    R(\varepsilon)<\frac{1}{a + \sqrt{\lambda_R}},
\end{equation}
where $\lambda_R$ is the largest positive eigenvalue of $-R_{0i0j}$ and $a$ denotes the maximum acceleration of the system along its motion. If the trapping potential which localizes the system is sufficiently strong, a good estimate for the localization of the system is $\sqrt{\langle\hat{\bm x}^2\rangle}$, so that the framework can be applied when $\sqrt{\langle\hat{\bm x}^2\rangle}<1/(a+\sqrt{\lambda_R})$. However, different systems might require different methods of estimating their localization depending on the specific shape of the trapping potential.

Another important assumption for our setup is that the energy of the system is non-relativistic (that is, sufficiently smaller than the system's rest energy). Although this assumption is not explicitly required in order to construct the formalism, it is important so that the prescribed Hamiltonian of Eq. \eqref{eq:Hcurved} is able to accurately predict the relativistic corrections to the internal dynamics of the system. In fact, in Subsection \ref{sub:coincidence} we show that under this assumption the framework of Subsection \ref{sub:main} yields the most relevant relativistic corrections for the description of wavefunctions in curved spacetimes found in~\cite{jonas}. Moreover, we remark again that other relativistic modifications to the prescription of Eq. \eqref{eq:Hcurved} might be required, which take into account the relationship of the internal degrees of freedom of the system with the geometry of spacetime and the trajectory's motion.

We now comment on the relationship between the non-relativistic quantum theory and the framework of general relativity. In our model, we assumed that spacetime is not affected by the quantum system. This assumption is reasonable provided that the stresses and energy of the quantum system are small enough, so that their effect on the spacetime metric can be neglected. This a usual assumption even when considering relativistic quantum theories in curved spacetimes, so that it is natural to expect that this assumption also has to be made in our treatment. Moreover, considering the effect of superpositions of quantum systems in a background spacetime has been argued to lead to superpositions of spacetimes~\cite{flaminiaSuperpositionSpacetimes2022}, which are not within the scope of this manuscript.

Finally, we comment on the relationship of our model with causality, and what is missed in this treatment. The formalism developed here allows one to consider the effect of the local geometry of a curved spacetime in the dynamics of a quantum system locally via the dependence on the position operator $\hat{\bm x}$. Although this interaction affects the system's position degree of freedom locally, in a non-relativistic setup this effect changes a single degree of freedom (the system's wavefunction) at each value of $\tau$. In essence, this means that the local effect of curvature in each portion of the system affect\tb{s} it instantaneously in its own frame. Essentially, information within this quantum system propagates acausally. This is expected from any non-relativistic quantum theory, and is also the case for our framework from Subsection \ref{sub:main}. The consequences of this causality violation have been carefully studied in the literature in similar setups~\cite{PipoFTL,BrunoDan}. An expected consequence of this violation is the ability for non-relativistic systems to signal between spacelike separated points, violating causality at the order of their effective size. We remark that no physical system should be able to signal faster than light, and this can be seen as another bound for the regime of validity of our framework, similar to what happens with specific non-relativistic quantum systems in curved spacetimes~\cite{us2}.

Overall we conclude that the framework presented in this manuscript can be employed to accurately describe non-relativistic quantum systems provided that 1) the system is sufficiently localized with respect to the Fermi bound, 2) its non-relativistic energy is sufficiently smaller than its rest energy, and 3) one is not interested in its uses in communication protocols between regions which are spacelike separated by proper distances which are of the order of the size of the system. 





\subsection{The coincidence limit and first order corrections}\label{sub:coincidence}

In this section we describe the first order corrections emerging due to motion of the system and spacetime curvature due to the addition of the redshift factor of Eq. \eqref{eq:redshift}.

In~\cite{jonas}, a non-relativistic quantum formalism has been presented for the approximate description of a localized fermionic particle in curved spacetimes. The authors then traced over the spin degrees of freedom of the particle, obtaining the free Hamiltonian for a localized non-relativistic wavefunction added to corrections in terms of curvature and acceleration. For systems such that their non-relativistic energy is much smaller than their rest mass, the most relevant correction terms found were $ma_i \hat{x}^i$ and $\frac{m}{2}R_{0i0j}\hat{x}^i\hat{x}^j$. \tb{These corrections were also found in~\cite{roura} and~\cite{theguy}.} We will now show that to first order in curvature and acceleration, these corrections are also the ones provided by the formalism presented in Subsection \ref{sub:main}. 

The new dynamics introduced by the formalism we presented are due to the introduction of the redshift factor in the original Hamiltonian which localizes the quantum system. Using the expansions of Eq. \eqref{eq:expansionFNC}, we can write the redshift factor as
\begin{equation}
    \gamma(\tau,\bm x) = 1 + a_i x^i + \frac{1}{2} R_{0i0j}x^ix^j + \mathcal{O}(r^3).
\end{equation}
Considering only the first order corrections in Eq. \eqref{eq:Hcurved} due to spacetime curvature and acceleration (which are also the corrections to second order in the system's localization), one obtains the approximate Hamiltonian
\begin{equation}\label{eq:Kwurtz}
    \hat{\mathcal{H}}(\tau) \approx \hat{H}(\tau) + \frac{1}{2}\left(\left(a_i \hat{x}^i + \tfrac{1}{2} R_{0i0j}\hat{x}^i\hat{x}^j\right)\hat{H}(\tau) + \text{H.c.}\right).
\end{equation}
Notice that the Hamiltonian $\hat{H}(\tau)$ can be split into two contributions: the rest mass of the system, and its non-relativistic energy, associated to an operator $\hat{H}_{\textsc{nr}}(\tau)$ which contains its kinetic and potential energies,
\begin{equation}
    \hat{H}(\tau)  = m +\hat{H}_{\textsc{nr}}(\tau).
\end{equation}
The assumption that the system's degrees of freedom are non-relativistic then implies that its non-relativistic energy is much smaller than its rest mass, $\langle\hat{H}_{\textsc{nr}}(\tau)\rangle\ll m$. In particular, this implies that  the terms of the form $\left(a_i \hat{x}^i + \tfrac{1}{2} R_{0i0j}\hat{x}^i\hat{x}^j\right)\hat{H}_{\textsc{nr}}(\tau)$ in Eq. \eqref{eq:Kwurtz} contribute much less than the terms which involve the rest mass of the system. In fact, reintroducing units of $c$, the term $a_i x^i$ picks up a factor of $1/c^2$, while the factor of $R_{0i0j}\hat{x}^i\hat{x}^j$ is of the order of the ratio of the size of the system by the curvature radius of spacetime. It is then safe to neglect these terms in most non-relativistic setups. 

Under these approximations, we can write
\begin{equation}
    \left(a_i \hat{x}^i + \tfrac{1}{2} R_{0i0j}\hat{x}^i\hat{x}^j\right)\!\hat{H}(\tau)\approx m\!\left(a_i \hat{x}^i + \tfrac{1}{2} R_{0i0j}\hat{x}^i\hat{x}^j\right),
\end{equation}
so that the Hamiltonian that promotes time evolution with respect to the time parameter $\tau$ can be written as
\begin{equation}\label{eq:approxH}
    \hat{\mathcal{H}}(\tau)\approx \hat{H}(\tau)+m  a_i(\tau) \hat{x}^i + \frac{m}{2}R_{0i0j}(\tau)\hat{x}^i \hat{x}^j.
\end{equation}
Notice that the correction terms arise from the product of the rest mass term $m\openone$ with the redshift factor. Due to the fact that the rest mass term in the Hamiltonian is proportional to the identity, and that $\gamma(\tau,\hat{\bm x})$ only depends on $\bm x$, the leading order relativistic corrections to the Hamiltonian are independent of the ordering ambiguities that might show up (See Eq. \eqref{eq:Hcurved} and related discussions). The Hamiltonian of Eq. \eqref{eq:approxH} also precisely matches the coupling of a localized system with curvature used in~\cite{gravDetec,pitelli} and the most relevant corrections found in~\cite{jonas}. However, in~\cite{jonas}, other corrections of the order of acceleration and curvature times the system's non-relativistic energy are also found. These extra corrections are not naturally taken into account in our formalism, and can be neglected if the non-relativistic rest energy of the system is sufficiently smaller than its rest mass. If, on the other hand, one wishes to consider these corrections, it is enough to add them to the prescription of the Hamiltonian in Eq. \eqref{eq:Hcurved}.

Eq. \eqref{eq:approxH} gives a Hamiltonian which is naturally self-adjoint under this non-relativistic approximation. Eq. \eqref{eq:approxH} also gives a simple expression for $\hat{\mathcal{H}}(\tau)$, with a quadratic correction to the Hamiltonian $\hat{H}(\tau)$. This approximation is valid provided that the energy of the system is much smaller than $mc^2$ and that the system is more localized than both the curvature radius of spacetime and than $c^2/a$, where we reintroduced the factors of $c$ in order to make the limits of validity more explicit.

\subsection{Example: a uniformly accelerated quantum harmonic oscillator in constant curvature spacetimes}\label{sub:example}

In this section we exemplify the formalism developed in this section to a harmonic oscillator with uniform proper acceleration in a constant curvature spacetime. For convenience, we will assume \tb{the spatial dimension to be} $n=3$ in this example.

The Riemann curvature tensor in a constant curvature spacetime takes the shape
\begin{equation}
    R_{\mu\nu\alpha\beta} = \alpha (g_{\mu\alpha}g_{\nu \beta} - g_{\mu\beta}g_{\nu\alpha}),
\end{equation}
where $\alpha$ is a constant, related to the Ricci curvature by $\alpha = R/12$. The sign of $R$ determines whether spacetime is positively ($R>0$) or negatively ($R<0$) curved, corresponding to deSitter and anti-deSitter spacetimes, respectively. In particular, in Fermi normal coordinates associated to a trajectory $\mf{z}(\tau)$, we obtain
\begin{equation}\label{eq:Rconst}
    R_{0i0j}(\mf z(\tau)) = -\alpha\delta_{ij},
\end{equation}
so that $R_{0i0j}\hat{x}^i \hat{x}^j = -\alpha \hat{\bm x}^2$, which is the relevant quantity for the redshift factor $\gamma(\tau,\hat{\bm x})$ that influences the Hamiltonian of the system.

The Hamiltonian for a quantum harmonic oscillator with frequency $\omega$ and mass $m$ can then be written as Eq. \eqref{eq:HO} by taking $V(\bm x) = m \omega^2 \bm x^2/2$. That is,
\begin{equation}
    \hat{H} = m + \frac{\hat{\bm p}^2}{2m} + \frac{m \omega^2}{2}\hat{\bm x}^2.
\end{equation}
In order to describe the system in a background curved spacetime, we must then have that the particle is localized in a region sufficiently smaller than the Fermi bound. A quantum harmonic oscillator is localized in a region of the order of $\sqrt{\langle\hat{\bm x}^2\rangle}=r/\sqrt{m\omega}$, where $r=\sqrt{2\langle\hat{n}\rangle+1}$. We then use the approximation derived in Appendix \ref{app:fermi} for the Fermi bound, $\ell \gtrsim 1/(a+\sqrt{\lambda_R})$, where $\lambda_R$ is the largest negative eigenvalue of $R_{0i0j}$. Using Eq. \eqref{eq:Rconst}, we find that $\lambda_R = \text{max}(0,-\alpha)$. Thus, the formalism developed in this section can be applied if the relevant states for the setup satisfy
\begin{equation}
    \frac{r}{\sqrt{m \omega}}<\frac{1}{a + \sqrt{\lambda_R}}.
\end{equation}
If the spacetime is positively curved, we can then describe the system if the frequency of the harmonic oscillator is larger than $a^2r^2/m$. If spacetime is negatively curved, curvature reduces the Fermi bound, so that one must have a frequency larger than $(a+ \sqrt{R/12})^2r^2/m$. In either case, we will assume that the gap of our harmonic oscillator is large enough so that we can employ our formalism.

Employing the approximation of Eq. \eqref{eq:approxH} described in Subsection \ref{sub:coincidence}, we then obtain the Hamiltonian of the system
\begin{equation}
    \hat{\mathcal{H}} = m + m\, \bm a \cdot \hat{\bm x} -\frac{m \alpha}{2} \hat{\bm x}^2 + \frac{\hat{\bm p}^2}{2m} + \frac{m\omega^2}{2}\hat{\bm x}^2.
\end{equation}
Given that all terms in the Hamiltonian above are quadratic, it still defines a quantum harmonic oscillator. In fact, $\hat{\mathcal{H}}$ can be rewritten as
\begin{equation}
    \hat{\mathcal{H}} = m -\frac{m \bm a^2}{2(\omega^2 - \alpha)}+ \frac{\hat{\bm p}^2}{2m} + \frac{m(\omega^2-\alpha)}{2}\left(\hat{\bm x}+\frac{\bm a}{\omega^2 - \alpha}\right)^2.
\end{equation}
From the equation above we can see that the resulting theory is a harmonic oscillator with frequency $\omega'$ given by $\omega'^2 = \omega^2 -\alpha$, with wavefunctions shifted by the vector $- \bm a /\omega'^2$. The acceleration of the system and curvature of spacetime also shift the energy of the ground state by $-m\bm a^2/2\omega'^2$.

Notice that if the curvature of spacetime is too large (with $R>0$), then it is possible that the effective potential generated by deSitter curvature is larger than the trapping potential of the oscillator, which would result in the particle accelerating away from the center of the coordinate system. In this regime, the wavefunction of the system cannot be considered to be localized, and the condition of Fermi localization breaks down, so that this formalism would not be suitable for its description. 

On the other hand, anti-deSitter spacetime creates an effective trapping potential for the particle, and increases the frequency of the oscillator. This phenomena has also been seen for a localized fermionic system in~\cite{jonas}, where it was found that the curvature of anti-deSitter spacetime can be responsible for trapping the particle even when $\omega = 0$.

\section{Application to Particle Detector Models}\label{sec:detectors}

In this section we relate the framework presented in Section \ref{sec:NRQS} to the formalism of particle detector models. We formulate a general notion of particle detector in Subsection \ref{sub:generalPDs}, and we analyze the consequences of the dynamics induced by the system's acceleration and the curvature of spacetime in Subsection \ref{sub:consequences}.

\subsection{General particle detector models}\label{sub:generalPDs}

Broadly speaking, a particle detector model is any localized non-relativistic quantum system that couples locally to a quantum field. The first instance of a particle detector was introduced by Unruh~\cite{Unruh1976}, where he considered a particle in a box linearly coupled to a scalar quantum field. This model was later simplified by DeWitt~\cite{DeWitt}, who considered a finite dimensional system for the detector's internal degrees of freedom. Over the past years more general particle detector models have been introduced in the literature, considering smeared detectors that could couple to different fields and operators in different quantum field theories~\cite{Takagi,Sachs1,sachs2018entanglement,eduAchimBosonFermion,Pozas-Kerstjens:2015,Nicho1,richard,mine}. In this section we will present a framework which generalizes these previous models.

In order to describe a particle detector model from the framework presented in Section \ref{sec:NRQS}, we require two extra ingredients apart from the non-relativistic quantum system: a quantum field theory and an interaction between the quantum field and the localized system. In this section we will refer to the non-relativistic quantum system as ``the detector''.

Consider a quantum field theory for a field $\hat{\phi}^a(\mf x)$ in the spacetime $\mathcal{M}$, where $a$ denotes any collection of Lorentz indices, associated with the spin of the field. We describe the quantum field theory as an association of smooth compactly supported fields $f_a(\mf x)$ to elements of a *-algebra $\mathcal{A}$. In general, one assumes $\mathcal{A}$ to satisfy important properties related to causality, commutation relations and conjugation, and a detailed description of these for fields of different spin can be found in~\cite{aqft}. For our purposes, one can think of a quantum field as an operator valued distribution $\hat{\phi}^a(\mf x)$ that acts in compactly supported fields $f_a(\mf x)$ according to
\begin{equation}
    \hat{\phi}(f) = \int \dd V\, \hat{\phi}^a(\mf x) f_a(\mf x).
\end{equation}
The derivatives and other operators acting in the field can then be obtained via functional differentiation and integration by parts. Regularized products of field operators at each point may then be introduced by extending the algebra $\mathcal{A}$ (see~\cite{Khavkine2015} for details). This extended set $\mathcal{A}$ then defines the set of valid operators in the quantum field theory.

The interaction of the detector with the field is prescribed in terms of an interaction Hamiltonian which couples to an operator-valued distribution of the quantum field theory, say $\hat{O}^b(\mf x)$, where $b$ stands for any collection of Lorentz indices (not necessarily the same as the field $\hat{\phi}^a(\mf x)$. In order to produce a scalar interaction Hamiltonian, one must have an operator in the detector's Hilbert space which is a tensor of the same rank as that of $\hat{O}^b(\mf x)$. We define the tensor operator as $\hat{\mu}(\tau) = \hat{\mu}^b(\tau)\mf{E}_b$, where $\hat{\mu}^b(\tau)$ is an operator in $\mathcal{H}_\textsc{x}^{(\tau)}\otimes\mathcal{H}_\textsc{s}$ and $\mf E_b$ denotes the orthonormal frame for tensors of the same rank as $\hat{O}^b(\mf x)$ built from the extended Fermi frame $\mf e_\mu$. We further assume $\hat{\mu}^b(\tau)$ to be only a function of the operators $\hat{\bm x}$, $\hat{\bm p}$ and $\{\hat{s}_i\}$ and of the time parameter $\tau$. For convenience, we will work in the interaction picture from now on, so that $\hat{\mu}^b(\tau)$ includes the free time evolution associated to the detector's ``free''\footnote{We use quotation marks for the word free because one usually picks the free Hamiltonian to be time independent, and that is not necessarily the case here. More discussion about this can be found in Subsection \ref{sub:consequences}.} Hamiltonian $\hat{\mathcal{H}}(\tau)$. Then, the interaction Hamiltonian is prescribed in the interaction picture as
\begin{equation}\label{eq:generalPD}
    \hat{H}_I(\tau) = \lambda \gamma(\tau,\hat{\bm x})\hat{\mu}^\dagger_b(\tau)\hat{O}^b(\tau,\hat{\bm x}) + \text{H.c.},
\end{equation}
where $\hat{O}^b(\tau,\tb{\bm x})$ denote\tb{s} the components of the operator $\hat{O}^b(\mf x)$ in the frame $\mf{E}_b$\footnote{We consider $\hat{O}^b(\mf x)$ and $\hat{\mu}^b(\tau)$ to be written in the orthonormal frame $\mf E_b$ in order to avoid unnecessary metric prescriptions in the contraction.} evaluated in Fermi normal coordinates around the curve $\mf z(\tau)$, \tb{ and the replacement of the dependence in the (classical) coordinates $\bm x$ by the quantum position operator $\hat{\bm x}$ formally means} 
\begin{equation}
    \tb{\hat{O}^b(\tau,\hat{\bm x}) \coloneqq  \int
\dd\Sigma \,\hat{O}^b (\tau , \bm{x}) \ket{\bm{x}}\!\!\bra{\bm{x}}_{\tau},}
\end{equation}
\tb{where the subscript $\tau$ in $\ket{\bm x}\!\!\bra{\bm x}_\tau$ denotes time evolution with respect to the detector's free Hamiltonian: $\ket{\bm x}\!\!\bra{\bm x}_\tau = \hat{U}^\dagger(\tau)\ket{\bm x}\!\!\bra{\bm x}\hat{U}(\tau)$ with $\hat{U} = \mathcal{T}\exp(-\ii \int^\tau \hat{\mathcal{H}}(\tau')\dd \tau')$.} In Eq. \eqref{eq:generalPD}, $\gamma(\tau,\hat{\bm x})$ denotes the redshift factor of Eq. \eqref{eq:redshift} and $\hat{\mu}_b^\dagger(\tau)$ denotes the dual field to $\hat{\mu}^b(\tau)$ (for instance, for a spinor field, $\hat{\mu}_b^\dagger(\tau)$ stands for $\hat{\bar{\mu}}{}^\dagger(\tau) = \gamma^0\hat{{\mu}}^\dagger(\tau)$). 

Common examples of particle detector models found in the literature can be recovered when the Hamiltonian $\hat{\mathcal{H}}(\hat{\bm x},\hat{\bm p},\{\hat{s}_i\},\tau)$ is independent of $\tau$ and $\hat{\mu}^b(\tau)$ can be written entirely in terms of the position operator, say $\hat{\mu}^b(\tau) = \mu^b(\tau,\hat{\bm x})$. For simplicity, we also assume that there are no internal degrees of freedom associated with $\mathcal{H}_\textsc{s}$. In this case, we can expand the Hamiltonian $\hat{H}_I(\tau)$ in terms of the position basis of the detector, $\ket{\bm x}$ via Eq. \eqref{eq:idDecomp}. We obtain
\begin{align}
    \hat{H}_I(\tau) &= \lambda \int \dd \Sigma \,\gamma(\tau,\bm x)  \mu_b^*(\tau,\bm x) \hat{O}^b(\tau,\bm x)\ket{\bm x}\!\!\bra{\bm x}_\tau +\text{H.c.}\nonumber\\
    &= \lambda \int \dd^n \bm x \sqrt{-g}\, \mu_b^*(\mf x) \hat{O}^b(\mf x)\ket{\bm x}\!\!\bra{\bm x}_\tau  +\text{H.c.},
\end{align}
where we wrote $\mf x = (\tau,\bm x)$ and used the fact that $\sqrt{g_\Sigma}\,\gamma(\mf x) = \sqrt{-g}$ in the second equality. We then assume the detector's free Hamiltonian $\hat{\mathcal{H}}$ to be independent of $\tau$ and to have discrete energy eigenvectors, $\ket{\psi_n}$ with energy eigenvalues $E_n$ with $\hat{\mathcal{H}}\ket{\psi_n} = E_n \ket{\psi_n}$. The eigenfunctions are defined as $\psi_n(\tau,\bm x) = \braket{\bm x}{\psi_n(\tau)} = e^{-\ii E_n \tau}\psi_n(\bm x)$, where we write the wavefunction at $\tau = 0$ as $\psi_n(\bm x)$. In the eigenbasis of the free Hamiltonian, the interaction Hamiltonian reads
\begin{align}
    \hat{H}_I(\tau) &= \!\lambda \sum_{nm}\!\int\! \dd^n \bm x \sqrt{-g}\, \psi_n^*(\bm x)\psi_m(\bm x)e^{i\Omega_{nm}\tau}\!\mu_b^*(\mf x) \hat{O}^b(\mf x)\nonumber\\
    &\,\,\,\,\,\,\,\,\,\,\,\,\,\,\,\,\,\,\,\,\,\,\,\,\,\,\,\,\,\,\,\,\,\,\,\,\,\,\,\,\,\,\,\,\,\,\,\,\,\,\,\,\,\,\,\,\,\,\,\,\,\,\,\,\,\times\ket{\psi_n}\!\!\bra{\psi_m}+\text{H.c.},
\end{align}
where $\Omega_{nm} = E_n - E_m$ is the energy gap between the states labelled by $n$ and $m$. Then, in order to draw a better comparison with the previous models in the literature (See e.g. \cite{us,us2,mine}), we define the spacetime smearing tensors $(\Lambda_{nm})^b(\mf x) = \psi_n(\bm x)\psi_m^*(\bm x)\mu^b(\mf x)$. We can then write the Interaction Hamiltonian as
\begin{align}\label{eq:Henergybasis}
    \hat{H}_I(\tau)\! &= \!\lambda\! \sum_{nm}\!\!\int \!\!\dd^n \bm x \sqrt{-g}\,(\Lambda_{nm})_b^*(\mf x) \hat{O}^b(\mf x)e^{\ii\Omega_{nm}\tau}\!\ket{\psi_n}\!\!\bra{\psi_m}\nonumber \\&\,\,\,\,\,\,\,\,\,\,\,\,\,\,\,\,\,\,\,\,\,\,\,\,\,\,\,\,\,\,\,\,\,\,\,\,\,\,\,\,\,\,\,\,\,\,\,\,\,\,\,\,\,\,\,\,\,\,\,\,\,\,\,\,\,\,\,\,\,\,\,\,\,\,\,\,\,\,\,\,\,\,\,\,\,\,\,\,\,\,\,\,\,\,\,+\text{H.c.}
\end{align}
The integrand in the expression above can be identified as the Hamiltonian density. In fact, most recent studies that consider finite-sized particle detectors in curved spacetimes (for instance~\cite{us,us2,PipoFTL,antiparticles,mine,ericksonBH,carol,geometry,ahmed,ericksonKen}) prescribe the interaction of the detector with the field in terms of the Hamiltonian density in order to highlight the locality of the theory. In the approach presented here, locality is implemented in terms of the dependence on the position operator of the non-relativistic quantum system.

Overall, the model of Eq. \eqref{eq:generalPD} for the interaction of a localized non-relativistic quantum system with a quantum field represents the most general interaction between a non-relativistic quantum system localized around a trajectory and an operator in a quantum field theory. The considerations about the description of non-relativistic quantum systems in curved spacetimes from Section \ref{sec:NRQS} (including its regimes of validity and covariance of the model) also apply to the general particle detector models presented here, and naturally impose a limit for the regime of validity for these models. 

Although the model of Eq. \eqref{eq:generalPD} is very general, and as we will see, can recover many models in the literature, it is not able to implement some features of specific models in the literature, specially when it comes to delocalization of the center of mass of detectors, which was considered in~\cite{achimDelocalized,achimDelocalizedHarvesting}, for instance. This delocalization would amount to describing the curve $\mf z(\tau)$ quantum mechanically, which would require significant changes in our formalism, and are out of the scope of this manuscript.

As we will show in the examples below, previous particle detector models used in the literature can be recovered from the model of Eq. \eqref{eq:generalPD} by choosing an appropriate quantum system, together with the detector and field operator that mediate the interaction.

\subsubsection{The scalar UDW model} 

The simplest scalar UDW model found in the literature consists of a two-level system coupled to a real scalar quantum field $\hat{\phi}(\mf x)$ according to the interaction Hamiltonian weight~\cite{us,us2,PipoFTL,BrunoDan,antiparticles}
\begin{equation}\label{eq:UDW}
    \hat{h}_I(\mf x) = \lambda \Lambda(\mf x) (e^{i \Omega \tau}\hat{\sigma}^++e^{-i \Omega \tau}\hat{\sigma}^-)\hat{\phi}(\mf x),
\end{equation}
where $\lambda$ is the coupling constant, $\Lambda(\mf x)$ is the (real) spacetime smearing function, $\Omega$ is the energy gap between the two-levels of the system, and $\hat{\sigma}^\pm$ are the raising and lowering operators.

The interaction of Eq. \eqref{eq:UDW} can be recovered from the general model of Eq. \eqref{eq:generalPD} from any non-relativistic quantum system by restricting it to two-levels and neglecting the terms of the interaction which commute with the detector's free Hamiltonian\footnote{\tb{T}he terms which commute with the free Hamiltonian would only promote a shift in the energy levels of the theory, and would not show up in the leading order corrections to most quantities of interest when considering particle detectors. For this reason, these terms can be ignored for most applications of particle detector models.}. Consider a localized quantum system that is entirely described by its position degrees of freedom and has ($\tau$ independent) discrete energy levels $E_n$ with eigenstates $\ket{\psi_n}$. We prescribe the interaction with the quantum field by the Hamiltonian
\begin{equation}\label{eq:intermediaryUDW}
    \hat{H}(\tau) = \lambda\gamma(\tau,\hat{\bm x}) g(\tau,\hat{\bm x}) \hat{\phi}(\tau,\hat{\bm x}),
\end{equation}
where $g(\mf x) = g(\tau,{\bm x})$ is any real scalar function evaluated in Fermi normal coordinates. We then identify the operator $\hat{\mu}(\tau) = g(\tau,\hat{\bm x})$ and the field operator \mbox{$\hat{O}(\mf x) = \hat{\phi}(\mf x)$}, so that Eq. \eqref{eq:intermediaryUDW} reads
\begin{align}
    \hat{H}_I(\tau) &= \lambda \sum_{nm}\int \dd^n \bm x \sqrt{-g}\,\Lambda_{mn}(\mf x) \hat{\phi}(\mf x)e^{i\Omega_{nm}\tau}\ket{\psi_n}\!\!\bra{\psi_m},
\end{align}
with $\Omega_{nm} = E_n - E_m$ and $\Lambda_{nm}(\mf x) = \psi_n^*(\bm x)\psi_m(\bm x) g(\tau,\bm x)$.

Restricting this system to two energy levels with an energy gap $\Omega$, say $n = g$ and $m = e$, we obtain the Hamiltonian
\begin{align}
    \hat{H}_I(\tau) = \lambda \int & \dd^n \bm x \sqrt{-g}\,\hat{\phi}(\mf x)\,\\
    &\Big(\Lambda_{eg}(\mf x)e^{\ii\Omega\tau}\ket{\psi_e}\!\!\bra{\psi_g} +\Lambda_{ge}(\mf x)e^{-\ii\Omega\tau} \ket{\psi_g}\!\!\bra{\psi_e}\nonumber\\
    &\:\:\:\:\:\:\:\:\:\:\:+\Lambda_{gg}(\mf x)\ket{\psi_g}\!\!\bra{\psi_g}+\Lambda_{ee}(\mf x)\ket{\psi_e}\!\!\bra{\psi_e}\Big),\nonumber
\end{align}
where we split the terms which do not commute with the free Hamiltonian in the second line, and the terms which do in the third line. As previously mentioned, in order to recover the UDW model, we neglect the terms which commute with the free Hamiltonian of the detector. This is a reasonable assumption if one is mostly interested in the excitation and de-excitation of the detector, as these terms do not contribute to energy level excitations to leading order in perturbation theory. In order to recover the UDW model, we assume that \mbox{$\psi_g(\bm x)\psi^*_e(\bm x) = \psi^*_g(\bm x)\psi_e(\bm x)$}, so that \mbox{$\Lambda(\mf x) = \Lambda_{ge}(\mf x) = \Lambda_{eg}(\mf x)$}, and the interaction Hamiltonian reads
\begin{align}
    \hat{H}_I(\tau) \!=& \lambda\!\! \int\!\! \dd^n \bm x \sqrt{-g} \Lambda(\mf x)\!\!\left(e^{\ii\Omega\tau}\!\!\:\!\ket{\psi_e}\!\!\bra{\psi_g} \!+\!e^{\!-\ii\Omega\tau} \!\:\!\!\ket{\psi_g}\!\!\bra{\psi_e}\right)\!\hat{\phi}(\mf x).
\end{align}
Denoting $\hat{\sigma}^+ = \ket{\psi_e}\!\!\bra{\psi_g}$ and $\hat{\sigma}^- = \ket{\psi_g}\!\!\bra{\psi_e}$, one then identifies the exact same Hamiltonian density from Eq. \eqref{eq:UDW}.

It is worth mentioning that instead of neglecting the part of the interaction Hamiltonian which commutes with the detector's free Hamiltonian, one could alternatively model an auxiliary internal degree of freedom for the detector in $\mathcal{H}_\textsc{s} = \mathbb{C}^2$ in order to recover the UDW model. Then, by adding an energy gap for states $\ket{0},\ket{1}\in\mathbb{C}^2$ one could choose the ground and excited states as $\ket{\psi_g'} = \ket{\psi_g,0}$ and $\ket{\psi_e'} = \ket{\psi_e,1}$ and \mbox{$\hat{\mu}(\tau,\hat{\bm x}) = g(\hat{\bm x})(\hat{\sigma}^+(\tau) + \hat{\sigma}^-(\tau))$}. These choices together with the reduction to the two dimensional subspace spanned by $\ket{\psi_g'}$ and $\ket{\psi_e'}$ reduce the model of Eq. \eqref{eq:generalPD} to Eq. \eqref{eq:UDW} exactly. Thus, we have argued how the general particle detector model from Eq. \eqref{eq:generalPD} can be used to recover the most used particle detector model in the literature by picking a quantum system which yields the corresponding spacetime smearing function $\Lambda(\mf x)$.

Finally, we mention that one could choose any other scalar operator for $\hat{O}(\mf x)$, such as $:\!\!\hat{\phi}^2(\mf x)\!\!:\,$, which would give detector models studied in~\cite{eduAchimBosonFermion,Sachs1,sachs2018entanglement}, for instance. Moreover, the quantum field theory that this detector couples to can be more general than a scalar field theory. For instance, for a spinor field $\hat{\psi}(\mf x)$, the operator $\hat{O}(\mf x)$ can be chosen as $\hat{O}(\mf x) \!= \,:\!\!\hat{\bar{\psi}}(\mf x) \hat{\psi}(\mf x)\!\!:\,$, which would recover other models studied in~\cite{eduAchimBosonFermion,Sachs1,sachs2018entanglement}, for instance. A generalization of the reduction presented here can also be carried naturally for the case of complex scalar fields, recovering the scalar models of~\cite{neutrinos,antiparticles,carol}. Overall, the model of Eq. \eqref{eq:generalPD} can be used to recover any coupling with a scalar quantum field, or with a scalar operator in a more general quantum field theory.




\subsubsection{The light-matter interaction} 

In order to recover models of particle detectors based on the light-matter interaction, and to extend these to curved spacetimes, we consider our non-relativistic quantum system to be a hydrogen atom. That is, its free Hamiltonian is prescribed as
\begin{equation}\label{eq:atom}
    \hat{H}_\textrm{atom} = \frac{\hat{\bm p}^2}{2 m_e}-\frac{e^2}{4\pi |\hat{\bm x}|},
\end{equation}
where $m_e$ is the reduced mass of the electron, and $e$ its fundamental charge. This Hamiltonian admits bound states $\ket{\psi_{nlm}}$ labelled by three quantum numbers, with $n\in \mathbb{N}$, $0\leq l \leq n-1$, $-l\leq m \leq l$, so that \mbox{$\hat{H}_\textrm{atom}\ket{\psi_{nlm}} = E_n \ket{\psi_{nlm}}$}, where $E_n = - \alpha^2 m_e/2n^2$ and $\alpha$ denotes the fine structure constant.

The interaction of an atom with a background quantum electromagnetic field then defines a particle detector model, where a localized non-relativistic quantum system (the atom) couples to a quantum field (the electromagnetic field). Although there are different ways of prescribing this interaction~\cite{Nicho1,richard}, here we will focus on the so-called dipole interaction, where, for inertial motion in flat spacetimes, the interaction Hamiltonian can be written as
\begin{equation}\label{eq:dipole}
    \hat{H}_I(t) = -e\,\hat{\bm x}\cdot \hat{\bm E}(t,\hat{\bm x}),
\end{equation}
where $\hat{\bm E}(t,\bm x)$ is the electric field in the frame of the atom. It is important to mention that the Hamiltonian of Eq. \eqref{eq:dipole} only accurately models the interaction of an atom with the electromagnetic field in some regimes\footnote{The interested reader can check~\cite{Nicho1,richard} for more detailed discussions.}. The Hamiltonians from Eqs. \eqref{eq:atom} and \eqref{eq:dipole}, together with a description for the electric field then determine a particle detector model for an inertial atom in flat spacetime.

In order to consider this system undergoing a trajectory $\mf z(\tau)$ in curved spacetimes, we must first check the conditions for the framework presented in Section \ref{sec:NRQS} to be applicable. That is, we must have that the system is more localized than the Fermi bound of the curve and that its non-relativistic energy is sufficiently smaller than its rest energy. For an electron in an atom, we have $|E|\sim 13,6\text{eV}$ and $m_e \approx 0.5\text{MeV}$, so that this approximation is valid. In order to address the localization of the atom, we use that an atom's extension in space can be approximately bounded by $1/|E|$, where $E$ is the average energy of the state considered. Assuming the Hydrogen atom to be in the state $\ket{\psi_{nlm}}$, and using the estimate for the Fermi bound of Eq. \eqref{eq:approximationFermi}, we find that a regime where the formalism can be used for this description is when
\begin{equation}
    \frac{1}{|E_n|} = \frac{2 n^2}{\alpha^2 m_e} < \frac{1}{a + \sqrt{\lambda_R}}.
\end{equation}
That is, an atom is well described undergoing an accelerated motion in curved spacetimes if its proper acceleration added to the square root of the largest eigenvalue of $-R_{0i0j}$ is smaller than $\alpha^2 m_e/2n^2$. For instance, for a hydrogen atom in its ground state undergoing uniformly accelerated motion in a flat spacetime, we obtain that its acceleration must be smaller than $10^{25}\text{m}/\text{s}^2$. This acceleration is large enough for probing an Unruh temperature as large as $10^6 \textrm{K}$, for instance.

Under the Fermi localization assumption, our framework then allows us to describe an atom undergoing a trajectory $\mf z(\tau)$ in a curved spacetime. We promote the $\hat{\bm x}$ and $\hat{\bm p}$ operators according to Eqs. \eqref{eq:posOp} and \eqref{eq:pOp}, and take into account the redshift factor discussed in Eq. \eqref{eq:redshift} in the Hamiltonian, which introduces new dynamics to the atom due to its acceleration and due to the curvature of spacetime. These can be approximated by Eq. \eqref{eq:approxH}.

In order to describe the coupling of the atom with the field, one must prescribe the interaction of Eq. \eqref{eq:dipole} covariantly, evaluating it in Fermi normal coordinates, and replacing $\bm x\longmapsto \hat{\bm x}$. We start by discussing the operator $\hat{\bm E}(\tau,\bm x)$. The effective electric field seen by an observer with four-velocity $v^\mu$ can be written as
\begin{equation}
    E^\mu(\mf x) = F^{\mu\nu}(\mf x)v_\nu.
\end{equation}
$E^\mu(\mf x)$ is then always a spacelike vector orthogonal to $v^\mu$, so that the electric field is defined in the rest space associated to the observer. The electric field associated to observers that move along constant $\bm x$ curves in Fermi normal coordinates can then be written as $E^\mu(\mf x) = F^{\mu\nu}(\mf x)u_\nu(\mf x)$, where $u^\mu(\mf x) = (\partial_\tau)^\mu/\sqrt{|g_{\tau\tau}|}$ is the four-velocity of observers moving along lines defined by $\bm x = \text{const}.$ The position vector $\bm x$ in the dipole interaction can then be associated with the position vector in Fermi normal coordinates, $\bm x = x^i \mf{e}_i$. Thus, the classical interaction Hamiltonian between the atom and the field can be written as
\begin{equation}
    {H}_I(\tau) = -e \gamma(\tau,{\bm x}){x}^i {E}_i(\tau,{\bm x}),
\end{equation}
where we added the classical redshift from Eq. \eqref{eq:redshift} and the contraction $x^i E_i$ naturally generalizes the dot product from Eq. \eqref{eq:dipole}. The quantum Hamiltonian then reads
\begin{equation}
    \hat{H}_I(\tau) = -e \gamma(\tau,\hat{\bm x})\hat{x}^\mu {E}_\mu(\tau,\hat{\bm x}),
\end{equation}
where $x^\mu = (0,x^i)$. Notice that this Hamiltonian is self-adjoint, because it is only a function of the position operator $\hat{\bm x}$. In terms of the $\hat{\mu}$ operator defined in Eq. \eqref{eq:generalPD}, this particle detector model couples to the electromagnetic field strength, $F_{\mu\nu}(\mf x)$, so that the $\hat{\mu}^{\mu\nu}(\tau)$ operator that defines it is \mbox{$\hat{\mu}^{\mu\nu}(\tau,\hat{\bm x}) = \hat{x}^\mu u^\nu(\tau,\hat{\bm x})$} and the coupling constant is given by $\lambda = - e$. Thus, we have shown how to describe the light-matter interaction in curved spacetimes as a particular case of the particle detector model of Eq. \eqref{eq:generalPD}.

\subsection{Consequences of the acceleration-curvature dynamics for particle detector models}\label{sub:consequences}

As discussed in Subsections \ref{sub:main} and \ref{sub:coincidence}, when one considers a non-relativistic quantum system undergoing a timelike trajectory in curved spacetimes, the acceleration of the trajectory and the curvature of spacetime influence the dynamics of the particle according the the redshift factor in Eq. \eqref{eq:redshift}. Under the assumptions that 1) the kinetic plus potential energy of the system is sufficiently small compared to its rest energy and 2) the system is sufficiently localized with respect to the curvature radius of spacetime and its inverse acceleration, the dynamics introduced by the system's motion and the background spacetime are associated with the Hamiltonian 
\begin{equation}\label{eq:Hrel}
    \hat{H}_{\text{rel}}(\tau) = m a_i(\tau) \hat{x}^i + \frac{m}{2}R_{0i0j}(\tau)\hat{x}^i \hat{x}^j
\end{equation}
from Eq. \eqref{eq:approxH}. These dynamics can by themselves promote energy level transitions within a particle detector, which generate an extra source of noise when probing a quantum field. That is, in order to be able to use a particle detector model to infer properties of a quantum field, one must make sure that the dynamics introduced by the system's trajectory and curvature of spacetime are sufficiently smaller than the effect of the interaction with the field. In this section, we discuss this regime in detail, and show how one could keep track of these new internal dynamics of particle detectors.

When one considers particle detector models, it is usual to assume that the Hamiltonian $\hat{H}$ in Eq. \eqref{eq:Hcurved} is time independent, so that one can apply perturbative techniques and use the interaction picture by picking the free Hamiltonian to be $\hat{H}$\footnote{Notice that while $\hat{H}$ is picked as the free Hamiltonian, the dynamics of the detector's internal degree of freedom are governed by the Hamiltonian $\hat{\mathcal{H}}(\tau) = \hat{H} + \hat{H}_{\text{rel}}(\tau)$.}. Notice that for general motion in a curved spacetime, the Hamiltonian $\hat{H}_\text{rel}(\tau)$ in Eq. \eqref{eq:Hrel} will depend on the time parameter $\tau$. In most cases, this forces one to treat the effect from the curve's acceleration and the curvature of spacetime perturbatively. Thus, in order to apply perturbative techniques and to use the interaction picture with free Hamiltonian $\hat{H}$, one would obtain an effective interaction Hamiltonian
\begin{equation}
    \hat{\mathcal{H}}_I(\tau) = \hat{H}_I(\tau) + \hat{H}_{\text{rel}}(\tau),
\end{equation}
where $\hat{H}_I(\tau)$ is the interaction with the quantum field from Eq. \eqref{eq:generalPD}. Thus, as expected, the resulting excitations of the detector will be both due to the interaction with the field and due to the relativistic corrections in $\hat{H}_\text{rel}(\tau)$.

The interaction unitary time evolution operator for the detector and field system can be written as
\begin{equation}\label{eq:UI}
    \hat{U}_I = \mathcal{T}_\tau\exp\left(-\ii \int \dd \tau  \hat{\mathcal{H}}_I(\tau) \right),
\end{equation}
where $\mathcal{T}_\tau$ denotes the time ordering operation with respect to the time parameter $\tau$. Notice that due to the non-relativistic nature of the detector, different choices of time parameter will in general result in different time evolution operators. This has been discussed in detail in~\cite{us2}, and the discussion naturally carries over to the general models presented here. In essence, this unitary time evolution operator promotes time evolution of the field and detector with respect to the proper time of the detector's trajectory. This is justified because the interaction with the field is also prescribed in this frame. 

The time evolution operator $\hat{U}_I$ is well defined under the assumption of Fermi localization. If the wavefunctions are not compactly supported, but instead are approximately Fermi localized, the effect of their tails is negligible, similar to the discussion in Subsection \ref{sub:regime}. Thus, within the regime of validity of the model, the operator $\hat{U}_I$ in Eq. \eqref{eq:UI} is well defined, and can be used to describe the unitary time evolution of the detector and field states.

Working perturbatively in the Hamiltonian $\hat{\mathcal{H}}_I(\tau)$, we can then apply the Dyson expansion to the time evolution operator, so that
\begin{equation}
    \hat{U}_I = \openone+\hat{U}_I^{(1)} + \hat{U}_I^{(2)}+\mathcal{O}(3), 
\end{equation}
where $\mathcal{O}(3)$ denotes terms of third order in $\hat{\mathcal{H}}_I(\tau)$ and
\begin{align}
    \hat{U}_I^{(1)} &= -\ii\int \dd \tau \hat{\mathcal{H}}_I(\tau),\\
    \hat{U}_I^{(2)} &= -\int \dd \tau\dd\tau' \hat{\mathcal{H}}_I(\tau)\hat{\mathcal{H}}_I(\tau')\theta(\tau-\tau'),
\end{align}
where $\theta(\tau-\tau')$ denotes the Heaviside step function, which introduces time ordering with respect to $\tau$. By writing $\hat{\mathcal{H}}_I(\tau) = \hat{{H}}_I(\tau)+\hat{{H}}_\text{rel}(\tau)$, we can split the unitary time evolution due to the interaction with the field ($\hat{{H}}_I(\tau)$) and due to the acceleration of the trajectory and curvature of spacetime ($\hat{{H}}_\text{rel}(\tau)$) as
\begin{align}
    \hat{U}_I^{(1)} &= \hat{U}_\text{field}^{(1)}+\hat{U}_\text{rel}^{(1)},\\
    \hat{U}_I^{(2)} &= \hat{U}_\text{field}^{(2)}+\hat{U}^{(2)}_{\text{mix}}+\hat{U}_\text{rel}^{(2)},
\end{align}
where
\begin{align}
    \hat{U}_\text{field}^{(1)}&= -\ii\int \dd\tau \hat{{H}}_I(\tau),\nonumber\\ \hat{U}_\text{rel}^{(1)}&= -\ii\int \dd\tau \hat{{H}}_\text{rel}(\tau)\nonumber,\\
    \hat{U}_\text{field}^{(2)}&= -\int \dd\tau\dd\tau' \hat{{H}}_I(\tau) \hat{{H}}_I(\tau')\theta(\tau-\tau'),\nonumber\\
    \hat{U}^{(2)}_{\text{mix}}\!&= \!-\!\!\int\!\! \dd\tau\dd\tau' \!\!\left(\!\hat{{H}}_I(\tau) \hat{{H}}_\text{rel}(\tau')\!+\!\! \hat{{H}}_\text{rel}(\tau)\hat{{H}}_I(\tau')\right)\!\theta(\tau\!-\!\tau')\nonumber,\\
    \hat{U}_\text{field}^{(2)}&= -\int \dd\tau\dd\tau' \hat{{H}}_\text{rel}(\tau) \hat{{H}}_\text{rel}(\tau')\theta(\tau-\tau').
\end{align}
We remark again that when considering particle detector models in the literature, one usually neglects the change in the internal dynamics of the detector due to its motion and the spacetime curvature. This corresponds to setting $\hat{H}_\text{rel} = 0$, so that the only contributions up to $\mathcal{O}(3)$ in the interaction are due to $\hat{U}_{\text{field}}^{(1)}$ and $\hat{U}_{\text{field}}^{(2)}$. The remaining terms are additional corrections predicted by our formalism.

In order to provide an explicit example, we consider the excitation probability for the detector to transition from a state $\ket{\psi_n}$ to an orthogonal state $\ket{\psi_m}$ when the field is in a Hadamard state $\omega$~\cite{kayWald,fewsterNecessityHadamard}, so that its odd \tb{n-}point functions vanish. Then, the excitation probability can be written as the expected value of the \emph{field operator}
\begin{equation}
    \bra{\psi_n}\hat{U}^\dagger_I\ket{\psi_m}\!\!\bra{\psi_m}\hat{U}_I\ket{\psi_n}
\end{equation}
in the state $\omega$. To second order in the interaction Hamiltonian, it is explicitly given by
\begin{equation}\label{eq:pnm}
    p_{n\rightarrow m} = p_{n\rightarrow m}^{\text{field}} + p_{n\rightarrow m}^{\text{rel}},
\end{equation}
where 
\begin{align}
    p_{n\rightarrow m}^{\text{field}} = &\,\lambda^2 \int \dd V\dd V'\\ &\!\!\!\!\times\Lambda_{nm}^b(\mf x)(\Lambda_{nm})^*_{b'}\!(\mf x')e^{-\ii \Omega_{nm}(\tau-\tau')}\!\langle\hat{O}^\dagger_b(\mf x)\hat{O}^{b'}(\mf x')\rangle_\omega\nonumber
\end{align}
is the excitation probability due to the interaction with the field, $\dd V = \sqrt{-g} \,\dd^{n+1}\mf x$ is the invariant spacetime volume element and $\Omega_{nm} = E_n - E_m$ is the energy gap between the states $\ket{\psi_n}$ and $\ket{\psi_m}$. The term $p_{n\rightarrow m}^{\text{rel}}$ in Eq. \eqref{eq:pnm} can be written as
\begin{align}
    p_{n\rightarrow m}^{\text{rel}} \!= \!\!\int\! \dd V \dd V' &\psi_n^*(\bm x)\psi_m(\bm x) \psi_n(\bm x')\psi_m(\bm x') e^{\!-\ii \Omega_{nm}(\tau-\tau')}\nonumber\\&\!\!\!\!\times\left(ma_i(\tau) x^i + \frac{m}{2}R_{0i0j}(\tau)x^i x^j\right),
\end{align}
and corresponds to the effect of the acceleration of the detector and the curvature of spacetime in the excitation probability. Thus, if the particle detector detector gets excited after its interaction, one can only claim that this excitation was due to the interaction with the field if $p_{n\rightarrow m}^{\text{rel}}\ll p_{n\rightarrow m}^{\text{field}}$.

At last, we comment on a scenario where $\hat{H}_{\text{rel}}$ is time independent, where the excitation of a detector can be completely associated with the quantum field, with a change to $\Omega_{nm}$ due to its acceleration and to the curvature of spacetime. Consider the situation where the spacetime is locally static, so that it possesses a locally defined timelike killing vector field $\xi$. Then, if the detector moves along of the flux of $\xi$ and its localization is within the region where $\xi$ is Killing, we have that $a_i = \text{const.}$ and $R_{0i0j} = \text{const.}$, so that $\hat{H}_\text{rel}$ is independent of $\tau$. This happens for a static detector in Schwarzschild spacetime, for instance. In this case, the Hamiltonian $\hat{\mathcal{H}}$ in Eq. \eqref{eq:approxH} is time independent, so that one can find time independent eigenstates $\ket{\psi_n'}$ and energy values $E_n'$ for $\hat{\mathcal{H}}$. Notice that $\ket{\psi_n'}$ and $E_n'$ will depend on $\hat{H}_\text{rel}$, and implicitly depend on $a_i$ and $R_{0i0j}$. In this case, the full $\hat{\mathcal{H}} = \hat{H} + \hat{H}_{\text{rel}}$ Hamiltonian can be considered the free Hamiltonian in the interaction picture, and one only obtains contributions to the time evolution operator arising from the interaction with the quantum field. This procedure can also be done for the example considered in Subsection \ref{sub:example}, with a uniformly accelerated harmonic oscillator in a constant curvature spacetime, where we saw that the energy gap gets shifted according to $\omega \longmapsto \sqrt{\omega^2 - R/12}$.

Overall, we conclude that one must be careful when using particle detector models in curved spacetimes in order to probe quantum fields, because a time varying proper acceleration and spacetime curvature may create an effective noise which blurs the effect of the field. However, if these terms are small enough compared to the effect of the quantum field in the detector, most of the excitation probability (and other observables of the theory) can be associated with the field. We also identified regimes where the detector's acceleration and curvature of spacetime can be implemented in the description of the detector, allowing one to associate detector excitations entirely with the quantum field.

\section{Conclusions}\label{sec:conclusions}

In this manuscript we provided a general recipe for describing a localized non-relativistic quantum system under going a timelike trajectory in curved spacetimes. Our framework can be applied to describe any system with one position degree of freedom which is sufficiently localized in space around the trajectory. If the system fulfills this condition, we showed how to consistently define position and momentum operators in the rest spaces of the trajectory, and identified the necessary relativistic modifications to the Hamiltonian in order to implement time evolution. 

We also defined the \emph{Fermi bound} $\ell$ associated to a given timelike trajectory in a curved background as the maximum proper length around the trajectory such that the Fermi normal coordinates can be used to fully describe a system with radius smaller than $\ell$. The Fermi bound then defines a size limit for quantum systems which can be described within our framework. Moreover, we estimated the Fermi bound in terms of the spacetime curvature and the proper acceleration of the curve, providing a lower bound for the regime of validity of the model.

Finally, we applied our description of non-relativistic quantum systems to particle detector models. This allowed us to define a general particle detector model: a sufficiently localized non-relativistic quantum system which couples to \emph{any} operator in a given quantum field theory. In particular, the model obtained can be reduced to other previously considered particle detector models studied in the literature, such as the UDW model and relativistic models for the light-matter interaction. We also identified the necessary modifications to the usual perturbative treatment of particle detectors, which take into account their motion and the curvature of spacetime. Overall this work is the first step towards a full characterization of localized non-relativistic quantum systems used to probe quantum fields in curved spacetimes.

\section*{Acknowledgements}

The author acknowledges Adam Teixid\'o-Bonfill for helping finding an estimate to the Fermi bound, Eduardo Mart\'in-Mart\'inez for discussions about the localization of non-relativistic quantum systems, Jos\'e Polo-Gomez for discussions regarding the definition of a consistent quantum theory, Bruno de S. L. Torres for discussions regarding the definition of the redshift factor associated to foliations in spacetime. The author thanks all members of the Barrio-RQI group (and Jos\'e de Ram\'on) for comments regarding the content of the manuscript. The author acknowledges support from the Natural Sciences and Engineering Research Council of Canada (NSERC) via the Vanier Canada Graduate Scholarship. Research at Perimeter Institute is supported in part by the Government of Canada through the Department of Innovation, Science and Industry Canada and by the Province of Ontario through the Ministry of Colleges and Universities. Finally, the author also thanks Profs. David Kubiz\v{n}\'ak and Eduardo Mart\'in-Mart\'inez’s funding through their NSERC Discovery grants.

\appendix

\section{Estimation of the Fermi bound}\label{app:fermi}

In this appendix we present an estimation for the Fermi bound, that is, we find an approximate bound for the maximum radius that a region contained in the rest spaces $\Sigma_\tau$ can have in order to be contained within the normal neighbourhood of the point $\mf z(\tau)$.

In order to obtain our estimate, we use the expansion of Eq. \eqref{eq:expansionFNC} for the
metric in Fermi normal coordinates. It is important to notice that the approximations of Eq. \eqref{eq:expansionFNC} are not only to second order in the distance of points to the curve, but are also the first order in curvature and acceleration expansions of the metric in Fermi normal coordinates. In the regime where these expansions are valid, the Fermi bound can be estimated as the maximum radius such that the metric does not become degenerate. That is, the largest radius such that the metric of Eq. \eqref{eq:expansionFNC}
is invertible. A good estimate for the $\tau$-Fermi bound can be given by considering the largest value of $|\bm x|$ such that 
\begin{equation}
    -(1+a_i x^i) - R_{0i0j}x^ix^j \neq 0.
\end{equation}
Notice that at $\bm x = 0$ this gives $-1$, so that for each $\tau$ we are looking for a bound on the radius $\ell_r$ such that 
\begin{equation}\label{eq:condition}
    |\bm x| \leq \ell_r \Rightarrow -(1+a_i x^i) - R_{0i0j}x^ix^j < 0. 
\end{equation}

In order to find $\ell_r$, let $a = \sqrt{a_ia^i}$ and define
\begin{equation}
    \lambda_R = \max_{|\bm x| = 1}\left(-R_{0i0j}x^ix^j\right).
\end{equation}
That is, $\lambda_R$ corresponds to the largest negative eigenvalue of $R_{0i0j}$, if there are any, else it is zero. We will now show that
\begin{equation}
    \ell_r = \frac{1}{a+\sqrt{\lambda_R}}
\end{equation}
fulfills the condition of Eq. \eqref{eq:condition}. Let $|\bm x|<\ell_r$, then
\begin{align}
    |\bm x|(a + \sqrt{\lambda_R})<1 \:\:\:\Rightarrow\:\:\: 1 - a|\bm x| >\sqrt{\lambda_R}|\bm x|,
\end{align}
so that using $a_i x^i \geq -a |\bm x|$,
\begin{equation}\label{eq:noname}
    1 + a_i x^i> \sqrt{\lambda_R} |\bm x| \:\:\:\Rightarrow\:\:\: (1 + a_i x^i)^2 > \lambda_R |\bm x| \geq - R_{0i0j}x^i x^j.
\end{equation}
This gives the desired result $0>-(1 + a_i x^i)^2- R_{0i0j}x^i x^j$.

We then obtain the approximate lower bound for the $\tau$-Fermi bound,
\begin{equation}\label{eq:tauFermiApprox}
    \ell_\tau \gtrsim \frac{1}{a+\sqrt{\lambda_R}},
\end{equation}
which is valid for points $\mf x$ such that $|\bm x|$ is sufficiently smaller than the curvature radius of spacetime and $1/a$, which is the regime of validity of the approximation of Eq. \eqref{eq:expansionFNC}. Notice that this bound works exactly in the case where there is no spacetime curvature, giving \mbox{$\ell_r = 1/a$}. The bound for the Fermi bound is then obtained by taking the infimum of Eq. \eqref{eq:tauFermiApprox} with respect to $\tau$.



\section{The redshift factor and the metric determinant}\label{app:det}

The metric in Fermi normal coordinates can be written as
\begin{equation}\label{eq:ADM}
    g = g_{\tau\tau}\dd\tau^2 +2g_{\tau i} \dd \tau \dd x^i+h_{ij}\dd x^i \dd x^j.
\end{equation}
Then, the inverse metric reads
\begin{equation}
    g^{-1} = g^{\tau\tau}\partial_\tau\otimes\partial_\tau + g^{\tau i}\partial_\tau \otimes \partial_i+g^{i\tau}\partial_i \otimes \partial_\tau + g^{ij}\partial_i\otimes\partial_j,
\end{equation}
where
\begin{align}
    g^{\tau\tau} &= \frac{1}{g_{\tau\tau} - g_{\tau i}g_{\tau j} h^{ij}},\\
    g^{\tau i} &= -\frac{h^{ij}g_{\tau j}}{g_{\tau\tau} - g_{\tau k}g_{\tau l} h^{kl}},\\
    g^{ij} &= h^{ij} + \frac{h^{ik}g_{\tau k}h^{jl}g_{\tau l}}{g_{\tau\tau} - g_{\tau k}g_{\tau l} h^{kl}},
\end{align}
and $h^{ij}$ is the inverse of the spatial metric, satisfying $h^{ik}h_{kj} = \delta^i_j$. The metric determinant can be written as
\begin{equation}
    \sqrt{-g} = \sqrt{h^{ij}g_{\tau i} g_{\tau j} -g_{\tau\tau}} \sqrt{g_{\Sigma}},
\end{equation}
where $g_{\Sigma} = \det(h_{ij})$.

The redshift factor associated to the foliation \mbox{$\tau = \text{const.}$} is determined by the norm of the 1-form $\dd \tau$, which is given by 
\begin{equation}
    {|\dd \tau|}^2 = g^{\tau\tau} = \frac{1}{g_{\tau\tau} - g_{\tau i}g_{\tau j} h^{ij}}.
\end{equation}
The redshift factor is then given by
\begin{equation}
    \gamma(\mf x) = \frac{1}{|\dd\tau|} = \abs{g_{\tau\tau} - g_{\tau i}g_{\tau j} h^{ij}}^\frac{1}{2},
\end{equation}
so that the invariant volume element of spacetime can be written as
\begin{equation}
    \dd V = \frac{\dd \tau}{|\dd \tau|}\wedge \dd \Sigma = \dd\tau \wedge(\gamma(\mf x) \dd \Sigma),
\end{equation}
where $\dd \Sigma = \sqrt{g_{\Sigma}}\,\dd^n \bm x$ is the measure in the \mbox{$\tau = \text{const}.$} surfaces.

\bibliography{references}
    
\end{document}